\documentclass[12pt, a4paper]{article}
\pdfoutput=1
\usepackage[utf8]{inputenc}
\usepackage[T1]{fontenc}
\usepackage{lmodern, textcomp}
\usepackage[american]{babel}
\usepackage{csquotes}
\usepackage{bbold}
\usepackage{multicol,caption}

\usepackage{mathtools}

\usepackage{eurosym}
\usepackage
[nottoc]{tocbibind}
\usepackage{authblk}
\usepackage{fancyhdr}
\usepackage{xspace}
\usepackage{bm}

\usepackage{graphicx}
\usepackage{subcaption}
\usepackage{float}
\usepackage{makecell}
\usepackage{multirow}
\usepackage{multicol} 
\usepackage{marginnote}
\usepackage[multiple]{footmisc}

\usepackage[usenames, dvipsnames, svgnames, table]{xcolor}

\usepackage[backend=bibtex8, citestyle=numeric-comp, maxbibnames=99,
			sorting=none, sortcase=false, sortcites=true, giveninits=true]{biblatex}

\usepackage{amsmath, amsfonts, amssymb}
\usepackage{mathtools}
\usepackage{mathrsfs}
\usepackage{cancel}
\usepackage{braket}
\usepackage{tensor}
\usepackage{slashed}
\usepackage{siunitx}
\usepackage{listings}

\usepackage{stmaryrd}
\usepackage{caption}
\usepackage{relsize,exscale}
\usepackage{array}
\usepackage[toc,page]{appendix}

\usepackage{enumerate}

\DeclareSymbolFont{largesymbols}{OMX}{cmex}{m}{n}

\setcellgapes{1pt}
\makegapedcells
\newcolumntype{R}[1]{>{\raggedleft\arraybackslash }b{#1}}
\newcolumntype{L}[1]{>{\raggedright\arraybackslash }b{#1}}
\newcolumntype{C}[1]{>{\centering\arraybackslash }b{#1}}

\newcommand{\Tr}{\mathrm{Tr}}

\newtheorem{proposition}{Proposition}
\newtheorem{remark}{Remark}

\newcommand{\beq}{\begin{equation}}
\newcommand{\eeq}{\end{equation}}
\newcommand{\bea}{\begin{eqnarray}}
\newcommand{\eea}{\end{eqnarray}}
\definecolor{mygray}{gray}{0.3}

\newcommand{\bes}{\begin{eqnarray}}
\newcommand{\ees}{\end{eqnarray}}

\newcommand\restr[2]{{
  \left.\kern-\nulldelimiterspace 
  #1 
  \vphantom{\big|} 
  \right|_{#2} 
  }}
\def\extd{\mathrm {d}}

\newcommand{\N}{\mathrm{N}}

\newcommand{\RE}{\mathrm{Re}}
\newcommand{\IM}{\mathrm{Im}}

\usepackage{multicol}
\usepackage{amssymb}

\def\XXint#1#2#3{{\setbox0=\hbox{$#1{#2#3}{\int}$}
     \vcenter{\hbox{$#2#3$}}\kern-.5\wd0}}

\usepackage[heightrounded, top=4cm, bottom=4cm, left=3cm, right=3cm, headheight=14pt]{geometry}

\usepackage[pdftex, breaklinks=true, linktocpage,
	colorlinks=true, urlcolor=blue, linkcolor=blue, citecolor=red
]{hyperref}

\newcommand{\email}[1]{\href{mailto:#1}{\nolinkurl{#1}}}
\newcommand{\emailfoot}[1]{\thanks{\email{#1}}}

\usepackage{marginnote}
\newcounter{draftcommentcnt}
\NewDocumentCommand{\draftcomment}{s O{red} m}{%
	\def\margnote{\IfBooleanTF{#1}{\marginnote}{\marginpar}}%
	\stepcounter{draftcommentcnt}%
	\textcolor{#2}{#3}%
	\margnote{\textcolor{#2}{$\Leftarrow$ \arabic{draftcommentcnt}}}%
}

\fancypagestyle{preprint}{
	\fancyhf{}

	\fancyhead[R]{\textsf{\preprint}}
}

\numberwithin{equation}{section}

\hypersetup{
	pdftitle={Functional renormalization group for $p=2$ matrix dynamics in the planar approximation},
}

\title{Functional renormalization group for ‘‘$p=2$'' like glassy matrices in the planar approximation\\
\bigskip
\Large{III. Equilibrium dynamics and beyond}}

\author[1]{Vincent Lahoche\emailfoot{vincent.lahoche@cea.fr}}
\author[1,2]{Dine Ousmane Samary\emailfoot{dine.ousmanesamary@cipma.uac.bj}}

\affil[1]{%
	Université Paris Saclay, \textsc{Cea}, \textsc{List}, Gif-sur-Yvette, F-91191, France
}
\affil[2]{%
	Faculté des Sciences et Techniques (ICMPA-UNESCO Chair)
	\protect\\
	Université d'Abomey-Calavi, 072 BP 50, Bénin
}
\begin{document}
\maketitle

\hrule
\hrule
\begin{abstract}
This paper is the last of the series investigating renormalization group aspects of stochastic random matrices, including a Wigner-like disorder. We consider the  equilibrium dynamics formalism that can be merged with the Ward identities arising from the large $\N$ effective kinetics. We construct a regulator that does not break time-reversal symmetry and show that the resulting flow equations reduce to the equilibrium flow built in our previous works. Finally, we investigate the flow equations beyond the equilibrium dynamics assumption and study the stability of the perturbation around the fluctuation-dissipation theorem.
\end{abstract}

\hrule
\hrule
\newpage
\pdfbookmark[1]{\contentsname}{toc}
\tableofcontents
\pagebreak

\section{Introduction}

In our two previous papers \cite{lahoche20241,lahoche20242}, we have considered a nonperturbative renormalization group for the equilibrium states corresponding to a stochastic complex $\N\times \N$ matrix $M$, characterized by a quenched disorder, materialized by a Wigner Hermitian matrix of size $\N$. In the large $\N$ limit, the equilibrium states look like a non-conventional Euclidean and non-local field theory, with an effective kinetics spectrum given by the Wigner law. We considered two different approximation schemes for solving the formal nonperturbative Wetterich equation. First in \cite{lahoche20241} we thought the standard vertex expansion at the leading order by using the derivative expansion, suitable for such a kind of non-local field theories \cite{benedetti2016functional,Lahoche_2017bb,Carrozza_2017a,Carrozza_2017}. This approach revealed some relevant fixed points in the deep IR, and strong evidence for the existence of a first-order phase transition. The existence of these interacting fixed points and this phase transition has to be confirmed in \cite{lahoche20242} where a method exploiting the Ward identities arising from effective kinetics has been used to close the hierarchy in the vicinity of local interactions. This method, inspiring from \cite{Lahoche:2018oeo,lahoche2022stochastic} allows us to go beyond the limitations of the vertex expansion, and in particular taking into account the intrinsic momenta dependency of effective vertices. From this method, we confirm the existence of a discontinuous phase transition, with the additional condition that anomalous dimensions vanish in the deep IR. Note that this condition agrees also with the vertex expansion, and we observed that the anomalous dimension goes to zero for some fixed point, as the order of the derivative expansion increases. Finally, the discontinuous phase transition we discovered can be characterized by the characteristics of the eigenvalue distribution of the Hermitian matrix $\chi=M^\dagger M$, and in particular by the larger edge bound of the distribution. 
\medskip

In the first part of this paper, we focus on the renormalization group for equilibrium dynamics of the stochastic matrices. In this regime, where the system is assumed to reach equilibrium with the time going to infinity, the fluctuation-dissipation theorem holds and should be taken into account in the construction of the truncation. This in particular imposes a constraint on the choice of the regulator \cite{duclut2017frequency,lahoche2021functional,lahoche2022functional}, which we trivially solve by imposing no coarse-graining in the frequency space. Solving Ward identities in that regime, we recover exactly the equation obtained in \cite{lahoche20242}, up to an irrelevant factor $\pi$ coming from the normalization of Fourier modes. In the second and last part of this paper, we investigate the out-of-equilibrium dynamics, a regime where the fluctuation-dissipation theorem breaks down. We then adapt the Ward identity method to this regime and deduce the flow equations in the IR limit. We then analyze the flow numerically and show the existence of a transition between in and out equilibrium phases, controlled by a fixed point whose we compute critical exponents. 
\medskip

The outline of the paper is the following. In sections \ref{sec1} and after a short presentation of the problem already detailed in \cite{lahoche20241,lahoche20242}, we construct the equilibrium dynamics formalism based on the Martin-Siggia-Rose formalism. In section \ref{sec2}, we construct the renormalization group solution in the deep IR using Ward identities, and in section \ref{sec3} we venture into the out-of-equilibrium dynamics. Finally, we summarize our results and some open issues in conclusion given in  section \ref{sec4}. Some additional materials are given in Appendices \ref{App1}, \ref{App2} (see also section \ref{App3}).

\section{Equilibrium dynamics formalism}\label{sec1}

In this section, we introduce the notations, and the basic formalism to construct a renormalization group in the equilibrium dynamics regime. In particular, we introduce the Martin-Siggia-Rose path integral, and a regularization scheme that preserves time-reversal symmetry, and especially its main incarnation: the fluctuation-dissipation theorem.

\subsection{The model}

Let us briefly provide  the definition of the model. We consider a complex stochastic matrix $M(t)\in \mathbb{C}^{\N\times \N}$, whose entries evolve accordingly with the stochastic process:
\begin{equation}
\frac{\extd M_{ij}}{\extd t}=-\frac{\delta \mathcal{H}}{\delta \overline{M}_{ij}(t)}+B_{ij}(t)\,,\label{eq1}
\end{equation}
where the noise $B(t)$ is a Gaussian matrix, with zero mean value and variance given by:
\begin{equation}
\langle \bar{B}_{i^\prime j^\prime}(t^\prime) B_{ij}(t) \rangle= T\delta_{ii^\prime}\delta_{jj^\prime}\delta(t-t^\prime)\,,
\end{equation}
and $\mathcal{H}$ reads as:
\begin{align}
\nonumber\mathcal{H}&:=\int_{-\infty}^{+\infty} \extd t\,\Tr \left( J M(t) M^\dagger(t) \right)+\Tr \left( K M^\dagger(t) M(t) \right)\\
&\quad+\sum_{p=1}^\infty\,\frac{a_p \N^{-p+1}}{(p!)^2} \,\int_{-\infty}^{+\infty} \extd t \,\Tr \, ({M}^\dagger(t)M(t))^p\,.
\end{align}
In the above relation 
 both $J$ and $K$ are Hermitian Wigner matrices \cite{potters2020first,forrester2010log} with the same variance $\sigma^2$. In the large $\N$ limit, the spectrum of the Wigner matrices becomes deterministic, given by the well-known semicircle law, meaning that for any suitable test function $f$, we have the following large $\N$-limit relation:
\begin{equation}
\frac{1}{\N}\lim_{\N \to \infty} \sum_{\mu=1}^\N\, f(\lambda_\mu)= \int_{-2\sigma}^{2\sigma} \, \extd \lambda \, \mu_W(\lambda) f(\lambda)\,, 
\end{equation}
with:
\begin{equation}
\mu_W(\lambda)=\frac{\sqrt{4\sigma^2-\lambda^2}}{2\pi \sigma^2}\,.
\end{equation}
Remark that the stochastic equation \eqref{eq1} admits a probabilistic interpretation.
We denote by $P(M,t)$ the probability distribution of  the random noise $\eta(t)$ which influences the trajectory such that $M(t)=M$, with the initial condition
 $M(t=0)=M_0$. Formally, for the complex matrices:
\begin{equation}
P(M,t):= \left\langle\prod_{i,j}\,  \delta(M_{ij}-M_{ij}(t)) \delta(\overline{M}_{ij}-\overline{M}_{ij}(t)) \right\rangle\,,\label{PMcomplex}
\end{equation}
and for Hermitian matrices:
\begin{align}
P(M,t):= \left\langle\prod_{i}\, \delta(M_{ii}-M_{ii}(t))\prod_{j<i} \delta(\RE{M}_{ij}-\RE{M}_{ij}(t)) \delta(\IM{M}_{ij}-\IM{M}_{ij}(t)) \rangle \right\rangle\,,\label{PMHermitian}
\end{align}
where $\RE(z)$ and $\IM(z)$ are the standard real and imaginary parts of $z$. The probability $P(M,t)$ follows a Fokker-Planck equation that can be easily deduced from the equation of motion \eqref{eq1}. For complex matrices, for instance, we have:
\begin{equation}
\frac{\partial P}{\partial t}= \mathbf{H} P[M,t]\,,\label{equEQ}
\end{equation}
where the Hamiltonian $\mathbf{H}$ is the second order derivative operator:
\begin{equation}
\mathbf{H}:= \sum_{i,j} \Bigg( T\frac{\partial^2}{\partial M_{ij} \partial \overline{M}_{ij}}+2\frac{\partial^2 H^{(\mathbb{C})}}{\partial M_{ij} \partial \overline{M}_{ij}}+\frac{\partial H^{(\mathbb{C})}}{\partial M_{ij}} \frac{\partial}{\partial \overline{M}_{ij}}+\frac{\partial H^{(\mathbb{C})}}{\partial \overline{M}_{ij}} \frac{\partial}{\partial {M}_{ij}}  \Bigg)\,.
\end{equation}
At equilibrium state, the large $\N$ partition function $Z_{\text{eq}}^\mathbb{C}[L,\overline{L}]$ reads setting $T=2$:
\begin{equation}
Z_{\text{eq}}^\mathbb{C}[L,\overline{L}]\underset{\N\to \infty}{\longrightarrow}\int \extd M \extd \overline{M}\, e^{- H_\infty^{\mathbb{C}}[M,\overline{M}]+ \overline{L}\cdot M+\overline{M}\cdot L}\,,\label{parteqcomplex2}
\end{equation}
where:
\begin{equation}
H_\infty^{\mathbb{C}}[M,\overline{M}]:= \sum_{\lambda,\mu} \overline{M}_{\lambda\mu} (\lambda+\mu+a_1){M}_{\lambda\mu}+\sum_{p=2}^\infty\,\frac{a_p \N^{-p+1}}{(p!)^2} \,\Tr \, ({M}^\dagger(t)M(t))^p\,,\label{HamiltonianCdef}
\end{equation}
and the dot product ‘‘$\cdot$'' is defined as:
$
A \cdot B:= \int \extd t\, \sum_{i,j=1}^N\, A_{ij}(t) B_{ij}(t)\,.
$
The resulting Boltzmann type partition function must be rewritten in a clever way introducing the two momenta, positives in the large $\N$:
\begin{equation}
p_1:= \lambda+2\sigma\,, \qquad p_2:= \mu+2\sigma\,,
\end{equation}
together with the mass: $m:=a_1-4\sigma$, such that the propagator becomes:
$
C_{p_1p_2}:=E^{-1}(p_1,p_2)\,,
$
where $E(p_1,p_2):=p_1+p_2+m$ is the \textit{energy} of the mode $(p_1,p_2)$. Then, in the large $N$ limit, the equilibrium partition function looks like (but differs from) an ordinary Euclidean field theory, with nearly continuous momenta $p$. 
\medskip

\subsection{Martin-Siggia-Rose path integral}

In our investigation, we only focus on the out-of-equilibrium dynamics that relax toward equilibrium for a long time. In this regime called \textit{equilibrium dynamics}, the Martin-Siggia-Rose formalism \cite{martin1973statistical,de2006random} allows expressing the Langevin dynamics \eqref{eq1} as a field theory. Once again, we recall the explicit construction for Hermitian matrices in the Appendix \ref{App1}, and in this section we only provide the results, assuming the reader is familiar with these concepts. 

\begin{proposition}
For complex random matrices, and for a given sample of the disorder matrices $J$ and $K$, the partition function $Z_{J,K}[L,\tilde{L}]$ describing field correlations in the equilibrium dynamics reads:
\begin{equation}
Z_{J,K}[L,\tilde{L}]:= \int \extd M \extd \overline{M} \extd \chi \extd \overline{\chi}\, e^{- S[M,\overline{M},\chi,\overline{\chi}]+\bar{L}\cdot M+\bar{M}\cdot L + \overline{\tilde{L}}\cdot \chi + \overline{\chi}\cdot \tilde{L}}\,,\label{outofeq}
\end{equation}
where $L$ and $\tilde{L}$ are source fields, and the MSR action is:
\begin{equation}
S[M,\bar{M},\chi,\overline{\chi}]:=\int_{-\infty}^{+\infty} \extd t\,\Tr\, \Big[T {\chi}^\dagger(t) \chi(t)+i{\chi}^\dagger(t) \big(\dot{M}+\partial_{\overline{M}} \mathcal{U}\big)+i  \big(\dot{{M}}^\dagger+\partial_{{M}^\tau} \mathcal{U}\big){\chi}(t) \Big]\,.\label{actionMSR}
\end{equation}
\end{proposition}
For a given sample of disorders $J$ and $K$, the functional $Z_{J,K}[L,\tilde{L}]$ generates correlation functions computed along a trajectory starting and ending at equilibrium, with boundary time $t=\pm \infty$. Note that the initial time $t_i$ can always be chosen as $t_i=-\infty$ because of the expected time translation invariance, and the end time $t_f$ to $+ \infty$ thanks to time-reversal symmetry. In other words, because of the relation assumption, \textbf{the system is expected to be arbitrarily close to equilibrium for any finite time} (see \cite{aron2010symmetries2} for more detail).
\medskip

\begin{remark}\label{remark2}
The definition of path integrals requires a choice for the discretization before constructing the continuum limit. The most popular choice in physics is the \textbf{Stratonovitch’s discretization prescription}, which consists of taking the value of the functions at the middle of the discrete step. With the \textbf{Îto prescription} in contrast, we took the value of the integrated functions on the discrete action at the left, on the boundary of the discretized step \cite{bouchaud1996mode,Zinn-Justin:1989rgp}. In this paper, we will use the Îto prescription, which in particular allows us to set to unity the determinant involved in the construction of the MSR path integral (see Appendix \ref{App1}). 
\end{remark}

The auxiliary field $\chi$ is a $\N \times \N$ matrix with time-dependent complex entries, and was introduced to break the square $\Tr \big(\dot{M}+\partial_{\overline{M}} \mathcal{U}\big)^\dagger \big(\dot{M}+\partial_{\overline{M}} \mathcal{U}\big)$, accordingly with the standard Hubbard-Stratonovich trick \cite{kleinert2011hubbard}. As recalled in the Appendix \ref{App1}, the MSR partition function is nothing but a clever way to rewrite the averaging of $\exp \left(\bar{L}\cdot M+\bar{M}\cdot L\right)$ over the realizations of the noise such that:
\begin{equation}
Z_{J,K}[L=0,\tilde{L}=0]=1\,,\label{normalization1}
\end{equation}
by construction. This property shows that the expectation function of any power of fields can be deduced by taking the functional derivative of the sources and setting $L=0,\tilde{L}=0$ at the end of the computation. The averaging of the correlation functions is furthermore obtained from the averaging of the partition function instead of the free energy and does not require replica \cite{castellani2005spin,de2006random}. Finally, due to the equilibrium dynamics assumption, the action \eqref{actionMSR} has to be invariant under time reversal, which corresponds to the field transformation:
\begin{equation}
M_{ij}^\prime(t)=M_{ij}(-t)\,,\quad \chi^\prime_{ij}(t)=\chi_{ij}(-t)+\frac{2i}{T} \dot{M}_{ij}(-t)\,, \label{transT1}
\end{equation}
and:
\begin{equation}
\bar{M}_{ij}^\prime(t)=\bar{M}_{ij}(-t)\,,\quad \bar{\chi}^\prime_{ij}(t)=\bar{\chi}_{ij}(-t)+\frac{2i}{T} \dot{\bar{M}}_{ij}(-t)\,. \label{transT2}
\end{equation}
The same consequence holds for the MSR path integral \eqref{actionMSR}, and the kinetic part of the action reads, in Fourier space:
\begin{align}
 \nonumber S_{\text{kin}}:=\int \extd \omega\,&\sum_{p_1,p_2}\, \Big[T\, \overline{\chi}_{p_1p_2}(\omega) \chi_{p_1p_2}(\omega)+i\,\overline{\chi}_{p_1,p_2}(\omega)\left(i\omega+E(p_1,p_2)\right) {M}_{p_1,p_2}(\omega)\\
&+i{\overline{M}}_{p_1,p_2}(\omega)\left(-i\omega+E(p_1,p_2)\right){\chi}_{p_1,p_2}(\omega) \Big]\,,\label{Skin}
\end{align}
where the Fourier transform is defined with the convention, for some function\footnote{Note that we use the same symbol $f$ for the function and its Fourier transform.} $f$:
\begin{equation}
f(\omega):=\frac{1}{\sqrt{2\pi}}\int_{-\infty}^{+\infty}\,\extd t\, f(t) e^{i\omega t}\,.
\end{equation}
The bare propagators read directly from \eqref{Skin}, 
\begin{equation}
C_{\bar{\chi}T}=\frac{{\omega}-i(p_1+p_2+m)}{{\omega}^2+(p_1+p_2+m)^2}\,,\quad C_{\bar{T}{\chi}}=-\frac{{\omega}+i(p_1+p_2+m)}{{\omega}^2+(p_1+p_2+m)^2}\,,\label{freepropa1}
\end{equation}
and
\begin{equation}
C_{T\bar{T}}=\frac{1}{T^2}\frac{1}{{\omega}^2+(p_1+p_2+m)^2}\,.\label{freepropa2}
\end{equation}
Furthermore, the response fields $\chi$ and $\bar{\chi}$ do not propagate:
\begin{equation}
C_{\chi\bar{\chi}}=0\,. \label{freepropaCHI}
\end{equation}
Interestingly, the result \eqref{freepropaCHI} valid at order zero in the perturbative expansion survives to all orders and may be considered as  an exact relation \cite{aron2010symmetries2}, meaning that component $\chi \bar{\chi}$ of the exact propagator $G$ (or equivalently the component $\bar{T} T$ of the mass matrix $\Gamma^{(2)}$) vanishes:
\begin{equation}
\boxed{
G_{\chi \bar{\chi}}(\omega,p_1,p_2)=0\,.} \label{conditionG}
\end{equation}
This is indeed the case for all higher correlation functions between response fields alone, see also the recent reference \cite{lahoche2022stochastic} for more detail:

\begin{remark}
All correlation functions implying the response field alone vanish:
\begin{equation}
\frac{\delta^{2n}Z_{J,K}}{\delta \chi^n \delta \overline{\chi}^n}\Big\vert_{L=0,\tilde{L}=0}=0\,.\label{conditiontrue}
\end{equation}
This can be proved by adding some linear Hamiltonian $\bar{M}\cdot k+\bar{k}\cdot M$ to the initial Hamiltonian. This is equivalent to translate the source fields as $\tilde{L}\to \tilde{L}-i k$ and $\bar{\tilde{L}}\to \bar{\tilde{L}}-i \bar{k}$, and \eqref{conditiontrue} follows from \eqref{normalization1}.
\end{remark}

\subsection{Interpolating classical and quantum action}

\begin{figure}
\begin{center}
\includegraphics[scale=0.9]{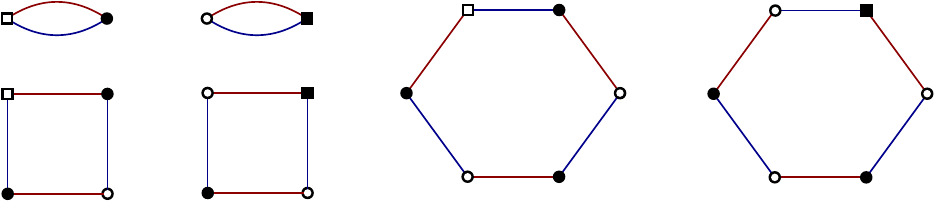}
\end{center}
\caption{Interactions bubbles of order $2$, $4$ and $6$ for the out of equilibrium theory.}\label{figgraphOUT}
\end{figure}

We consider the Wetterich formalism, aiming to interpolate between the MSR action and the effective action $\Gamma$, including all the quantum effects. 
We start with an explicit expression of the action in the Fourier representation. Usually, the interpolation is given by adding a regulator $R_k$, transforming the kinetic action (with the regulator term) of the MSR action as (see \eqref{Skin} and \cite{lahoche2022stochastic}):

\begin{align}
 \nonumber S_{\text{kin}}:=&\int \extd \omega\,\sum_{p_1,p_2}\, \Big[\, \overline{\chi}_{p_1p_2}(\omega) \chi_{p_1p_2}(\omega)+i\,\overline{\chi}_{p_1,p_2}(\omega)\left(i\omega+E(p_1,p_2)\right) {M}_{p_1,p_2}(\omega)\\\nonumber
&+i\,{\overline{M}}_{p_1,p_2}(\omega)\left(-i\omega+E(p_1,p_2)\right){\chi}_{p_1,p_2}(\omega)+i\,\overline{\chi}_{p_1,p_2}(\omega) R_k(p_1,p_2) {M}_{p_1,p_2}(\omega)\\
&+i\,{\overline{M}}_{p_1,p_2}(\omega)R_k(p_1,p_2){\chi}_{p_1,p_2}(\omega)\Big]\,.\label{actionMSR2}
\end{align}
Note that, we set $T=1$, which is always allowed up to a suitable rescaling of the fields and couplings involved in the MSR action. 
We denote as $\Xi:=(M,\chi)$ the doublet of fields, such that the regulator term $\Delta S_k$ reads also:
\begin{equation}
\Delta S_k=i \int \extd \omega\,\sum_{p_1,p_2}\, \bar{\Xi}_{p_1p_2}(\omega) R_k(p_1,p_2) {\Xi}_{p_1p_2}(\omega)\,,
\end{equation}
where the momentum-dependent mass $R_k(p_1,p_2)$, usually called \textit{regulator} ensures that IR contributions to the effective action are suppressed \cite{Delamotte_2012}. More precisely, we require that:
\begin{equation}
R_{k=0}(p_1,p_1)=0\,,\qquad R_{k\to 4\sigma}(p_1,p_2)\to \infty\,.
\end{equation}
As in our previous works \cite{lahoche20241,lahoche20242}, we choose:
\begin{equation}
R_k(p_1,p_2)=Z(k)k\left(\frac{4\sigma}{4\sigma-k}\right)\left(2-\frac{p_1+p_2}{k}\right) f\left(\frac{p_1}{k}\right)f\left(\frac{p_2}{k}\right)\label{eqR}
\end{equation}
where the function $f(x)$ is related to the Heaviside function:
$
f(x):=\theta(1-x),
$
such that, for $k$ small enough:
\begin{equation}
R_k(p_1,p_2)\approx Z(k)(2k-p_1-p_2)\theta(k-p_1)\theta(k-p_2)\,.
\end{equation}
We also define the quantity $\tilde{Z}:=4 \sigma/(4\sigma-k)$ introduced in \cite{lahoche20241}, which is equivalent to $Z$ in the deep IR. 
\medskip
The flow equation can be deduced, as usual, by taking the derivative of the partition function \cite{Delamotte_2012}, and the Wetterich equation takes the form:
\begin{equation}
\dot{\Gamma}_k=\int \extd \omega\,\sum_{p_1,p_2} \, \frac{\dot{R}_k(p_1,p_2)}{\Gamma_k^{(2)}(p_1,p_2)+R_k(p_1,p_2)}\Bigg\vert_{\omega,\omega}\,,\label{Wett2}
\end{equation}
where the notation $X\vert_{\omega,\omega}$ means that we only take into account the diagonal parts of the $2$-points function which is proportional to $\delta(0)$, because of the momenta conservation and is compensated by the same (infinite) factor on the left. As usual in out-of-equilibrium field theory based on the MSR formalism, we assume that  the interactions are linear in the response field $(\chi,\bar{\chi})$ \cite{lahoche2022stochastic,duclut2017frequency}. In the bubble graph representation, the interactions can then be represented by tripartite colored regular graphs, with black and white dots materializing fields $M, \bar{M}$ and black and white squares materializing (respectively) response field $\chi,\bar{\chi}$, with only one of such a square node per local components (with respect to the time). Figure \ref{figgraphOUT} gives some examples of the diagram representation of such interactions and the black and white dots and squares materialize respectively fields $M$ and $\eta$, and their complexes conjugate. In contrast, the colored edges materialize indices contractions. See also Appendices \ref{App2} and \ref{App3} below. 

\subsection{Local truncation and fluctuation-dissipation theorem}\label{secfloweq}

These defined given in the last section allow us to  introduce the theory space on which  the  interaction part of the effective average action $\Gamma_k$ reads:
\begin{equation}
\Gamma_{k,\text{int}}[\eta,\bar{\eta},\Phi,\bar{\Phi}]:= \sum_n i\frac{(1+\delta_{1n})u_{2n}}{n!(n-1)!} \sum_{p=1}^K\, \left(\overbrace{\vcenter{\hbox{\includegraphics[scale=0.65]{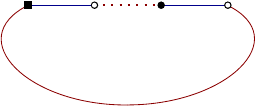}}}}^{\text{order} \,2p}\,+\,\overbrace{\vcenter{\hbox{\includegraphics[scale=0.65]{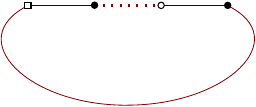}}}}^{\text{order}\, 2p}\right)\,,\label{expansionGamma}
\end{equation}
where $2K$ denotes the larger order interaction involved in the truncation. Note that we included the mass term (for $p=1$) in the interactions, with a factor $2$ in front. Furthermore, note that the definition of the bubbles includes integration over time, or frequency, depending if we work in direct or Fourier representation. For instance, explicitly:
\begin{equation}
\vcenter{\hbox{\includegraphics[scale=0.6]{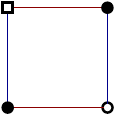}}}\,=\,\int_{-\infty}^{+\infty} \prod_{i=1}^4 \extd \omega_i\, \delta \left(\sum_i^4 (-1)^i\omega_i\right)\sum_{p_1,p_2,p_3,p_4} \bar{\eta}_{p_1p_2}(\omega_1)\Phi_{p_3p_2}(\omega_2) \bar{\Phi}_{p_3p_4}(\omega_3)\Phi_{p_1p_4}(\omega_4)
\end{equation}
For the kinetic truncation, we focus on the leading order of the derivative expansion, and choose:
\begin{align}
\nonumber \Gamma_{k,\text{kin}}[\eta,\bar{\eta},\Phi,\bar{\Phi}]&:= \int \extd \omega\,\sum_{p_1,p_2}\, \Big[\, \overline{\eta}_{p_1p_2}(\omega) \eta_{p_1p_2}(\omega)\\\nonumber
&+i\,\overline{\eta}_{p_1,p_2}(\omega)\left(i\omega+Z(k)(p_1+p_2)\right) {\Phi}_{p_1,p_2}(\omega)\\
&+i\,{\overline{\Phi}}_{p_1,p_2}(\omega)\left(-i\omega+Z(k)(p_1+p_2)\right){\eta}_{p_1,p_2}(\omega)\Big]\,.\label{kinetictruncationout}
\end{align}
This truncation is \textit{minimal} i.e. we do not consider field strength renormalization for response fields. Furthermore, the time-reversal symmetry given by the transformations \eqref{transT1} and \eqref{transT2} ensures that the wave function renormalization for response field must be equal to the renormalization of the terms $i \omega \bar{\eta} \Phi$ and $-i\omega \bar{\Phi} \eta$. More precisely, if $i \omega \bar{\eta} \Phi \to i \omega X_1(k)\bar{\eta} \Phi$, $-i\omega \bar{\Phi} \eta \to -iX_2(k)\omega \bar{\Phi} \eta$ and $\bar{\eta} \eta \to Y(k) \bar{\eta}\eta$, the time-reversal symmetry given by transformations \eqref{transT1} and \eqref{transT2} imposes:
\begin{equation}
X_1(k)=X_2(k)=Y(k)\,.
\end{equation}
For the same reason, the truncation should not include terms like $\bar{\Phi}\Phi$, and the expansion \eqref{kinetictruncationout} is the minimal one, compatible with time-reversal symmetry and causality -- see \cite{lahoche2022stochastic,duclut2017frequency,canet2011general} and Appendix \ref{App2} which involves the same  proof with the absence of a term involving only fields $M$ and $\bar{M}$, without response field. Accordingly, with \cite{lahoche2022stochastic}, we call \textit{heteroclicity} this property of interaction which necessarily includes the response field. In section \eqref{App3}, we propose a same  proof, based on Ward identities, explaining why the component $\bar{\Phi}\Phi$ of $\Gamma^{(2)}$ has to vanish. 

\begin{remark}\label{remarkPhiPhi}
Note that the absence of a $\bar{\Phi}\Phi$ term in the kinetic truncation can be easily traced from the condition that the response field does not propagate, i.e. the component $G_{k,\bar{\eta}\eta}$ of the effective propagator vanishes identically -- see for instance \cite{lahoche2022stochastic,aron2010symmetries2} and below.
\end{remark}
As illustration of the above prescription, we compute the flow equation for $u_2$ in the  expansion \eqref{expansionGamma} for $\Gamma_k$. Taking the second derivative of the Wetterich equation \eqref{Wett2} with respect to $\bar{\eta}$ and $\Phi$, we get, graphically:
\begin{equation}
\dot{\Gamma}_{k,\bar{\eta}\Phi}(p_1,p_2,\omega)\delta(\omega-\omega^\prime)=-\,\vcenter{\hbox{\includegraphics[scale=0.8]{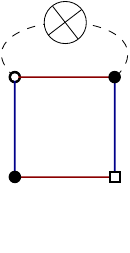}}}\,-\,\vcenter{\hbox{\includegraphics[scale=0.8]{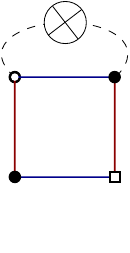}}}\,.\label{floweq2ptsOOE}
\end{equation}
 The crossed circle denotes the regulator insertion and dashed edges denote the effective propagator. As explained before, the effective propagator looks like a $2\times 2$ matrix in the doublet space $(\eta,\Phi)$. More precisely, we can write explicitly the $2\times 2$ matrix $\Gamma_k^{(2)}$ as:
\begin{equation}
\Gamma_k^{(2)}(p_1,p_2,\omega)=\begin{pmatrix}
1 & -\omega+iZ(k)(p_1+p_2+2k\tilde{u}_2) \\
\omega+iZ(k)(p_1+p_2+2k\tilde{u}_2)& 0
\end{pmatrix}
\,.\label{matrixGamma2}
\end{equation}
where we displayed columns and rows such that the diagonal involves $\bar{\eta}\eta$ components comes in first position at the top left. Also, we use the correspondence $u_2:=Zk \tilde{u}_2$. Note that the well-defined  two point function involves a product of Kronecker and Dirac delta $\delta(\omega-\omega^\prime)\delta_{p_1p_1^\prime}\delta_{p_2p_2^\prime}$ for conservation of momenta, that we implicitly canceled on both sides of the above equation. Adding the regulator contribution, the kinetic kernel can be easily inverted and the components of the propagator are:
\begin{equation}
G_{\bar{\eta}\Phi}(p_1,p_2,\omega)=\frac{1}{\omega+i Z f_k(p_1,p_2)}\,,\qquad G_{\bar{\Phi}\eta}(p_1,p_2,\omega)=\frac{1}{-\omega+i Z f_k(p_1,p_2)}\,,\label{effectivepropaout}
\end{equation}
and:
\begin{equation}
G_{\bar{\Phi}\Phi}(p_1,p_2,\omega)=\frac{1}{\omega^2+Z^2 f_k^2(p_1,p_2)}\,,
\end{equation}
where, 
\begin{equation}
f_k(p_1,p_2):=p_1+p_2+2k\tilde{u}_2+r_k(p_1,p_2)\,,\qquad R_k(p_1,p_2)=:Z(k)r_k(p_1,p_2)\,.
\end{equation}
These function in particular satisfy the \textit{fluctuation-dissipation theorem} \cite{aron2010symmetries2,lahoche2022stochastic}:
\begin{equation}
G_{\bar{\eta}\Phi}(p_1,p_2,\omega)-G_{\bar{\eta}\Phi}(p_1,p_2,-\omega)=2\omega G_{\bar{\Phi}\Phi}(p_1,p_2,\omega)\label{fluctuationdissipationTH}
\end{equation}
meaning that the choice of the regulator and truncation is compatible with the equilibrium dynamics assumption. Note that this is expected because the truncation agrees with the time reflection symmetry. Furthermore, this holds for any choice of the regulator, as soon as it remains independent.\footnote{In this case, the regulator is constrained by the time-reversal symmetry constraint, see \cite{lahoche2021functional,duclut2017frequency}.}
\medskip

Let us compute the flow for the mass $u_2$, which will be useful in the next section. To this end, we focus on one of the previous diagrams, for vanishing external momenta. The diagrams involve the loop is explicitly:
\begin{equation}
\mathcal{A}_2:=\int \, \extd \omega \, \int \, \rho(p) \extd p \frac{\dot{R}_k(p,0)}{\omega^2+Z^2f^2_k(p,0)}\frac{1}{\omega+i Z f_k(p,0)}\,,\label{defA2}
\end{equation}
which can be easily computed using the residue theorem ($A>0$):
\begin{equation}
\int \, \extd \omega \, \frac{1}{\omega^2+A^2}\frac{1}{\omega+i A}=-\frac{i\pi}{2A^2}\,,\label{integralkeyfrequencies}
\end{equation}
and we get:
\begin{equation}
\mathcal{A}_2:=\frac{-i}{8Z(k)k}\, \frac{J_{{\nu}}(k)}{(1+\tilde{u}_2)^2}\,,\quad \nu:=k \frac{\extd }{\extd k} \ln Z\,\label{A2result}
\end{equation}
and in the deep IR (see \cite{lahoche20241}):
\begin{equation}
J_{{\nu}}(k) \approx \frac{1}{\tilde{\sigma}^{3/2}}\,\left(\frac{14\nu(k)}{15}+\frac{7}{3}\right)\,,
\end{equation}
where:
\begin{equation}
\boxed{\tilde{\sigma}:=k^{-1} \sigma\,.}
\end{equation}
It is easy to check that the contributions of the two diagrams in \eqref{floweq2ptsOOE} are equal for vanishing external momenta. Furthermore, the contribution has to be counted twice, because of the two components $\bar{\eta} R_k \Phi$ and $\bar{\Phi} R_k \eta$. Finally, 
\begin{equation}
\dot{u}_2=+i u_4 \mathcal{A}_2(k)=-\frac{u_4}{8Z(k)k}\, \frac{J_{{\nu}}(k)}{(1+\bar{u}_2)^2}\,,
\end{equation}
and using dimensionless renormalized couplings, $\tilde{u}_2$, we recover essentially the equilibrium equation (up to a factor $\pi$):
\begin{equation}
\dot{\tilde{u}}_2=-(1+\nu)\tilde{u}_2-\frac{\tilde{u}_4}{8}\, \frac{J_{{\nu}}(k)}{(1+\tilde{u}_2)^2}\,.\label{flowequ2OOF}
\end{equation}
One can easily check that this holds for higher order equations and that up to a global rescaling of couplings (for $n>1$) by $\pi$, we recover the equilibrium flow equations derived in \cite{lahoche20241}. As explained in the introduction, this is expected because of the equilibrium dynamics assumption. However, if the equations are the same, the interpretation of the resulting phase space is different. In particular, regions where the equilibrium potential is unbounded from below, for instance, are interpreted as a region where the corresponding equilibrium state is non normalizable, and then where the system remains out of equilibrium for a long time. 

\section{Local truncation and Ward identities method}\label{sec2}

This section considers the Ward identity method introduced in \cite{lahoche20242}  for the equilibrium dynamics formalism. We begin with a general derivation of the Ward identities, arising because of the explicit $(\mathcal{U}(\N))^{\times 2}$ symmetry breaking of the effective kinetics term. We then use the equilibrium dynamics assumptions and show that the Ward identities in this case are reduced to their expression for the equilibrium theory -- up to the factor $\pi$ noticed before. 

\subsection{Modified Ward identities}

The Ward identities come from the underlying unitary symmetry, broken by the effective kinetic emerging at large $\N$ from the disorder contribution. The variation of MSR partition function \eqref{outofeq} with kinetic action modified by the regulator (see \eqref{actionMSR2}) leads to the functional identity:
\begin{align}
\nonumber &0=\int \extd\omega \Bigg[ i \delta E_k(p_1,p_1^\prime,p_2) \left( \frac{\delta^2}{\delta \tilde{L}_{\vec{p}\,}(\omega) \delta \bar{L}_{\vec{p}\,^\prime}(\omega)} +  \frac{\delta^2}{\delta \bar{\tilde{L}}_{\vec{p}\,^\prime}(\omega) \delta {L}_{\vec{p}\,}(\omega)}     \right)\\
& +\left( \bar{L}_{\vec{p}}(\omega)\frac{\delta}{\delta \bar{{L}}_{\vec{p}\,^\prime}(\omega)} -  {L}_{\vec{p}\,^\prime}(\omega)\frac{\delta}{\delta {{L}}_{\vec{p}}(\omega)}+ \bar{\tilde{L}}_{\vec{p}}(\omega)\frac{\delta}{\delta \bar{\tilde{L}}_{\vec{p}\,^\prime}(\omega)} -  \tilde{L}_{\vec{p}\,^\prime}(\omega)\frac{\delta}{\delta {\tilde{L}}_{\vec{p}}(\omega)}\right)\Bigg] e^{W_k(L,\tilde{L})}
\end{align}
where the free energy $W_k(L,\tilde{L})$ fluctuates via the index $k$, and
\begin{equation}
\delta E_k(p_1,p_1^\prime,p_2)= E_k(p_1^\prime,p_2)-E_k(p_1,p_2)\,, \quad E_k(p_1^\prime,p_2):= \vert \vec{p} \,\vert+m+R_k(p_1,p_2)\,.
\end{equation}
See the reference \cite{lahoche2022stochastic} for more detail. As for the equilibrium case, expressing Ward's identities in terms of observables is useful. We define connected correlation functions as:
\begin{equation}
G^{(n+\bar{n};m+\bar{m})}_{k, \vec{p}_1,\cdots,\vec{p}_n,\vec{\bar{p}}_1,\cdots,\vec{\bar{p}}_{\bar{n}},\vec{q}_1,\cdots,\vec{q}_m,\vec{\bar{q}}_1,\cdots,\vec{\bar{q}}_{\bar{m}}}=\prod_{i=1}^n \frac{\delta}{\delta L_{\vec{p}_i}} \prod_{j=1}^{\bar{n}} \frac{\delta}{\delta \bar{L}_{\vec{p}_j}} \prod_{I=1}^m \frac{\delta}{\delta \tilde{L}_{\vec{p}_I}} \prod_{J=1}^{\bar{m}} \frac{\delta}{\delta \bar{\tilde{L}}_{\vec{p}_J}} \, W_k(L,\tilde{L})\,.\label{correlationsDEF}
\end{equation}
This definition will be used when the explicit index structure is required. The classical fields as the functional derivatives of the free energy are then defined as:
\begin{equation}
\Phi_{\vec{p}}:=\frac{\delta W_k}{\delta \bar{L}_{\vec{p}}}\,,\qquad \bar{\Phi}_{\vec{p}}:=\frac{\delta W_k}{\delta {L}_{\vec{p}}}\,,\qquad \eta_{\vec{p}}:=\frac{\delta W_k}{\delta \bar{\tilde{L}}_{\vec{p}}}\,,\qquad \bar{\eta}_{\vec{p}}:=\frac{\delta W_k}{\delta {\tilde{L}}_{\vec{p}}}\,,
\end{equation}
and we get finally the following statement:
\begin{proposition}
The correlation functions of the equilibrium dynamical model satisfied the following Ward identity:
\begin{align}
\nonumber &\int \extd\omega \sum_{p_2} \Bigg[ i \delta E_k(p_1,p_1^\prime,p_2) \left( G_{k,\bar{\eta} \Phi}(\vec{p}\,^\prime,\omega; \vec{p},\omega)+\bar{\eta}_{\vec{p}}\, \Phi_{\vec{p}\,^\prime}
+G_{k,\bar{\Phi} \eta}(\vec{p},\omega; \vec{p}\,^\prime,\omega)+{\eta}_{\vec{p}\,^\prime}\, \bar{\Phi}_{\vec{p}} \right)\\
& +\left(\bar{L}_{\vec{p}}(\omega)\Phi_{\vec{p}\,^\prime}(\omega)-  {L}_{\vec{p}\,^\prime}(\omega) \bar{\Phi}_{\vec{p}\,}(\omega)+ \bar{\tilde{L}}_{\vec{p}}(\omega) \eta_{\vec{p}\,^\prime}(\omega)-  \tilde{L}_{\vec{p}\,^\prime}(\omega) \bar{\eta}_{\vec{p}}(\omega)\right)\Bigg]=0\,,\label{WardIDoutofEq2}
\end{align}
where , $\vec{p}:=(p_1,p_2)$ and $\vec{p}\,^\prime=(p_1^\prime,p_2)$. 
\end{proposition}

\subsection{Another ‘‘proof'' of heteroclicity}\label{App3}
Before coming to the main issue of this section, let us show that the Ward identities allow us to give another proof of heteroclicity i.e. a property that tells us why the effective action $\Gamma_k$ cannot include interactions independent of the response field (see Appendix \ref{App2} or \cite{lahoche2022stochastic} for more detail). 
 Remark also that the relations between observables arising from the unitary symmetry of interactions automatically consider the causality of the theory. Then we start by performing in the symmetric phase (the odd vertex function vanishes),  the second derivative of the Ward identity \eqref{WardIDoutofEq2} with respect to $\delta^2/\delta \Phi_{\vec{q}\,}(\Omega)\delta \bar{\Phi}_{\vec{\bar{q}}\,}(\bar{\Omega})$ and get
\begin{align}
\nonumber \int \extd\omega &\sum_{p_2} \Bigg[ i \delta E_k(p_1,p_1^\prime,p_2) \left( \frac{\delta^2G_{k,\bar{\eta} \Phi}(\vec{p}\,^\prime,\omega; \vec{p},\omega)}{\delta \Phi_{\vec{q}}(\Omega)\delta \bar{\Phi}_{\vec{\bar{q}}}(\bar{\Omega})}
+\frac{\delta^2 G_{k,\bar{\Phi} \eta}(\vec{p},\omega; \vec{p}\,^\prime,\omega)}{\delta \Phi_{\vec{q}}(\Omega)\delta \bar{\Phi}_{\vec{\bar{q}}}(\bar{\Omega})} \right)\\
& +\left(\frac{\delta\bar{L}_{\vec{p}\,}(\omega)}{\delta \bar{\Phi}_{\vec{\bar{q}}}(\bar{\Omega})}\delta_{\vec{p}\,^\prime \vec{q}\,}\delta(\omega-\Omega)-  \frac{\delta{L}_{\vec{p}\,^\prime}(\omega)}{\delta \Phi_{\vec{q}}(\Omega)} \delta_{\vec{p} \vec{\bar{q}}\,}\delta(\omega-\bar{\Omega})\right)\Bigg]=0\,.\label{Wardhetero1}
\end{align}
The two last terms are nothing but the components $\bar{\Phi}\Phi$ of $\Gamma_k^{(2)}$, which has to vanish because of the condition $G_{k,\bar{\eta}\eta}=0$\footnote{ We omit the bold notation for momenta on the right-hand side to improve readability.} (see Remark \ref{remarkPhiPhi}):

\begin{align}
\nonumber\frac{\partial^2G_{k,\bar{\eta}\Phi}(\vec{p}\,^\prime,{\omega};\vec{p},{\omega})}{\partial \Phi_{\bm q}({\omega}_1)\partial \bar{\Phi}_{\bm{\bar{q}}}({\bar{\omega}}_1)}=&- \sum_{p_1,p_1^\prime} G_{k,\bar{\eta}M}(p^\prime,p_1^\prime) \Gamma_{k,\Phi\bar{\Phi}\Phi\bar{\sigma}}^{(4)}(q,\bar{q},p_1^\prime,p_1) G_{k,\bar{\eta} \Phi}(p_1,p)\\
&-\sum_{p_1,p_1^\prime} G_{k,\bar{\eta}\Phi}(p^\prime,p_1^\prime) \Gamma_{k,\Phi\bar{\Phi}\Phi \bar{\Phi}}^{(4)}(q,\bar{q},p_1^\prime,p_1) G_{k,\bar{\Phi} \Phi}(p_1,p)\,,
\end{align}
and
\begin{align}
\nonumber\frac{\partial^2G_{k,\bar{\Phi}\eta}( \vec{p},{\omega};\vec{p}\,^\prime,{\omega})}{\partial \Phi_{\bm q}({\omega}_1)\partial \bar{\Phi}_{\bm{\bar{q}}}({\bar{\omega}}_1)}=&- \sum_{p_1,p_1^\prime} G_{k,\bar{\Phi}\eta}(p,p_1) \Gamma_{k,\Phi\bar{\Phi}\eta \bar{\Phi}}^{(4)}(q,\bar{q},p_1^\prime,p_1) G_{k,\bar{\Phi} \eta}(p_1^\prime,p^\prime)\\
&-\sum_{p_1,p_1^\prime} G_{k,\bar{\Phi}\Phi}(p,p_1) \Gamma_{k,\Phi\bar{\Phi}\Phi \bar{\Phi}}^{(4)}(q,\bar{q},p_1^\prime,p_1) G_{k,\bar{\Phi} \eta}(p_1^\prime,p^\prime)\,.
\end{align}
Note that to simplify the notation, we omitted the dependency on the external frequencies of $\Gamma^{(4)}_k$ (the vertex is assumed local to frequencies). We also use a graphical notation, as we introduced in our previous work \cite{lahoche20242}. For the component $\Gamma_{k,\Phi\bar{\Phi}\Phi \bar{\Phi}}^{(4)}$, we keep the same notation, and we denote as $\gamma_k^{(4)}$ the kernel vertex function, taking into account momenta conservation along external faces due to the connectivity of the effective planar functions. In the same way for components $\Gamma_{k,\Phi\bar{\Phi}\Phi\bar{\eta}}^{(4)}$ and $\Gamma_{k,\Phi\bar{\Phi}\eta\bar{\Phi}}^{(4)}$, we denote these kernels respectively as $\gamma_k^{(2+\bar{1},\bar{1})}$ and $\gamma_k^{(1+\bar{2},1)}$, accordingly with definition \eqref{correlationsDEF}. Note however that the contributions of heteroclites vertices vanish because of causality. Indeed, we expect from this causality behavior that $\langle \bar{\eta}_{\vec{p}}(t) M_{\vec{p}}(t^\prime) \rangle\propto \theta(t^\prime-t)$.  
Hence, the loop involves $\theta(0)$, which is zero because of the Îto convention (see Appendix \ref{App1}).  Graphically, the Ward identity reduces to:
\begin{equation}
\vcenter{\hbox{\includegraphics[scale=1]{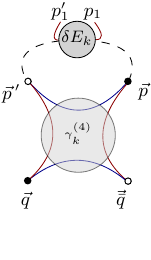}}}\quad + \quad \vcenter{\hbox{\includegraphics[scale=1]{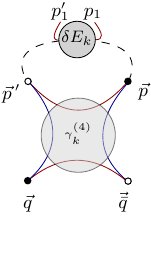}}}\,=\,0\,,
\end{equation}
whereas the gray disk materializes effective vertices and colored path materialize external faces. A direct inspection shows that this relation must be satisfied only if $\gamma_k^{(4)}=0$. The same thing can be shown recursively for higher couplings, then, because of causality, Ward identities directly imply heteroclicity. 

\subsection{Closing hierarchy: equilibrium dynamics}

Let us move on to the closing hierarchy issue. We derive the Ward identity \eqref{WardIDoutofEq2} with respect to $\delta^2/\delta \Phi_{\vec{q}\,}(\Omega)\delta \bar{\eta}_{\vec{\bar{q}}\,}(\bar{\Omega})$ and get in the symmetric phase:
\begin{align}
\nonumber \int \extd\omega \sum_{p_2} \Bigg[ &i \delta E_k(p_1,p_1^\prime,p_2) \Bigg( \frac{\delta^2G_{k,\bar{\eta} \Phi}(\vec{p}\,^\prime,\omega; \vec{p},\omega)}{\delta \Phi_{\vec{q}}(\Omega)\delta \bar{\eta}_{\vec{\bar{q}}}(\bar{\Omega})}
+\frac{\delta^2 G_{k,\bar{\Phi} \eta}(\vec{p},\omega; \vec{p}\,^\prime,\omega)}{\delta \Phi_{\vec{q}}(\Omega)\delta \bar{\eta}_{\vec{\bar{q}}}(\bar{\Omega})} \\\nonumber
&+\delta_{\vec{p}\,\vec{\bar{q}}}\,\delta_{\vec{p}\,^\prime\,\vec{q}}\,\delta(\omega-\Omega)\delta(\omega-\bar{\Omega})\Bigg) +\Bigg(\frac{\delta\bar{L}_{\vec{p}\,}(\omega)}{\delta \bar{\eta}_{\vec{\bar{q}}\,}(\bar{\Omega})}\delta_{\vec{p}\,^\prime \vec{q}\,}\delta(\omega-\Omega)\\
&-  \frac{\delta{\tilde{L}}_{\vec{p}\,^\prime}(\omega)}{\delta \Phi_{\vec{q}}(\Omega)} \delta_{\vec{p}\, \vec{\bar{q}}\,}\,\delta(\omega-\bar{\Omega})\Bigg)\Bigg]=0\,.\label{Wardclosing2}
\end{align}
The two last terms can be easily computed, because of the definition of the Legendre transform\footnote{Note that, as previously, we assume the effective vertices are local with respect to the frequency, and we discard Dirac delta and explicit frequency dependency to simplify the notations.}:
\begin{equation}
\frac{\delta\bar{L}_{\vec{p}\,}(\omega)}{\delta \bar{\eta}_{\vec{\bar{q}}}(\bar{\Omega})}= \left(\Gamma^{(2)}_{k,\bar{\eta}\Phi}(\vec{p}\,)+i R_k(\vec{p}\,) \right)\delta_{\vec{p}\,\vec{\bar{q}}}\,,
\end{equation}
and:
\begin{equation}
\frac{\delta{\tilde{L}}_{\vec{p}\,^\prime}(\omega)}{\delta \Phi_{\vec{q}}(\Omega)}=\left(\Gamma^{(2)}_{k,\bar{\eta}\Phi}(\vec{p}\,^\prime)+i R_k(\vec{p}\,^\prime\,) \right)\delta_{\vec{p}\,^\prime\,\vec{{q}}}\,.
\end{equation}
In the same way, taking into account heterocility and the fact that $G_{k,\bar{\eta}\eta}=0$:
\begin{equation}
\frac{\partial^2G_{k,\bar{\eta}\Phi}^{(\bar{1};1)}(\vec{p}\,^\prime,{\omega};\vec{p},{\omega})}{\partial \Phi_{\bm q}({\omega}_1)\partial \bar{\eta}_{\bm{\bar{q}}}({\bar{\omega}}_1)}=- \sum_{p_1,p_1^\prime} G_{k,\bar{\eta}\Phi}(p^\prime,p_1^\prime) \Gamma_{k,\Phi\bar{\Phi}\Phi\bar{\eta}}^{(4)}(q,\bar{q},p_1^\prime,p_1) G_{k,\bar{\Phi} \Phi}(p_1,p)\,,\label{decompositionG1}
\end{equation}
and
\begin{equation}
\frac{\partial^2G_{k,\bar{\Phi}\eta}^{(1;\bar{1})}(\vec{p}\,^\prime,{\omega};\vec{p},{\omega})}{\partial \Phi_{\bm q}({\omega}_1)\partial \bar{\eta}_{\bm{\bar{q}}}({\bar{\omega}}_1)}=- \sum_{p_1,p_1^\prime} G_{k,\bar{\Phi}\Phi}(p^\prime,p_1^\prime) \Gamma_{k,\Phi\bar{\Phi}\Phi\bar{\eta}}^{(4)}(q,\bar{q},p_1^\prime,p_1) G_{k,\bar{\Phi} \eta}(p_1,p)\,.\label{decompositionG2}
\end{equation}
Then, because two points functions are diagonal in the symmetric phase,
\begin{equation}
p_1=\bar{q}_1\,,\quad p_1^\prime=q_1\,,\quad q_2=\bar{q}_2,.\label{momentaconfig4ptsBIS}
\end{equation}
Defining:
$
\Gamma^{(2)}_{k,\bar{\eta}\Phi}(\vec{p}\,)=\omega+i \vert \vec{p}\,\vert-i\Sigma(\vec{p}\,)\,,
$
the Ward identity reads:
\begin{align}
\nonumber 2\sum_{p_2}\int \extd \omega\Big[&1+\frac{R_k(p_1^\prime,p_2)-R_k(p_1,p_2)}{p_1^\prime-p_1}\Big]G_{k,\bar{\Phi}\Phi}(\vec{p}\,,\omega)\,\gamma_{k,p_2,q_2,p_1^\prime,p_1}^{(2+\bar{1},\bar{1})}\,G_{k,\bar{\eta}\Phi}(\vec{p}\,^\prime,\omega)-\\
&\frac{\Sigma_k(\vec{p}\,^\prime)- \Sigma_k(\vec{p}\,)}{p_1^\prime-p_1}\Bigg\vert_{p_2=q_2} =0\,,
\end{align}
where the factor $2$ is a formal short notation for the two loop contributions having the same limit as $p_1\to p_1^\prime$. Then, taking the limit $p_1\to p_1^\prime \to 0$, we get:
\begin{equation}
\boxed{2\N\int \rho(q) \left[1+\frac{\extd R_k}{\extd p_1}(0,q)\right]\int \extd \omega G_{k,\bar{\Phi}\Phi}(0,q,\omega) G_{k,\bar{\eta}\Phi}(0,q,\omega)\gamma_{k,q000}^{(2+\bar{1},\bar{1})}+(\tilde{Z}(k)-1)=0\,.}\label{FMWI0MOUT}
\end{equation}
First, we remark that, for the term involving regulator, the same approximation as we used in the flow equation must hold, because the selected windows of momenta are the same. Therefore, it should make sense to replace propagators and vertex functions with their expression coming from the truncation (equations \eqref{kinetictruncationout} and \eqref{expansionGamma}) -- at least, this do not add any additional approximation with respect to the computation of the flow equation within this truncation. Then, we get for the term involving the regulator (we use again \eqref{integralkeyfrequencies} to compute the integral over frequencies):
\begin{align}
\nonumber\int \rho(q)& \frac{\extd R_k}{\extd p_1}(0,q)\int \extd \omega G_{k,\bar{\Phi}\Phi}(0,q,\omega) G_{k,\bar{\eta}\Phi}(0,q,\omega)\gamma_{k,q000}^{(2+\bar{1},\bar{1})}\\\nonumber
&=-\frac{i \pi}{2 \tilde{Z}}\gamma_{k,0000}^{(2+\bar{1},\bar{1})} \times \int \rho(q)\,\frac{\extd R_k}{\extd p_1}(0,q) \frac{1}{(f_k(0,q))^2}\\
&= -\frac{u_4}{\N\tilde{Z}}\frac{\left(\frac{1}{\tilde{\sigma }}\right)^{3/2}}{12 k^2}\frac{1}{(1+\tilde{u}_2)^2}\,.\label{equationtermcool}
\end{align}
In the above integral the first term denotes by $J(k)$ may be computed easily:
\begin{equation}
J(k):=2\N\int \rho(q) \int \extd \omega G_{k,\bar{\Phi}\Phi}(0,q,\omega) G_{k,\bar{\eta}\Phi}(0,q,\omega)\gamma_{k,q000}^{(2+\bar{1},\bar{1})}\,.
\end{equation}
To this ends, observes that for $s:=\ln (k)$:
\begin{equation}
\frac{\extd}{\extd s}\,G_{k,\bar{\Phi}\Phi}=-G_{k,\bar{\Phi}\bullet} \frac{\extd \Gamma_{k,\bullet\circ}^{(2)}}{\extd s} G_{k,\circ \Phi}-G_{k,\bar{\Phi}\bullet} \frac{\extd R_{k,\bullet\circ}}{\extd s} G_{k,\circ \Phi}\,,
\end{equation}
where in this formal expression, repeated symbols $\bullet:=(\Phi,\eta)$ and $\circ=(\bar{\Phi},\bar{\eta})$ are assumed to be summed and are considered as momenta, which are implicit in this expression.  Remark that the term with $(\bullet,\circ)=(\eta,\bar{\eta})$  vanished because of the flow equation, as soon as we assume the absence of effective vertices involving more than one $\eta$ component \cite{canet2011general}. The remaining terms are then:
\begin{align}
\nonumber \frac{\extd}{\extd s}\,G_{k,\bar{\Phi}\Phi}=&-G_{k,\bar{\Phi}\eta} \frac{\extd \Gamma_{k,\eta \bar{\Phi}}^{(2)}}{\extd s} G_{k,\bar{\Phi} \Phi}-G_{k,\bar{\Phi}\eta} \frac{\extd R_{k,\eta\bar{\Phi}}}{\extd s} G_{k,\bar{\Phi} \Phi}\\
& - G_{k,\bar{\Phi}\Phi} \frac{\extd \Gamma_{k,\Phi \bar{\eta}}^{(2)}}{\extd s} G_{k,\bar{\eta} \Phi}-G_{k,\bar{\Phi}\Phi} \frac{\extd R_{k,\Phi\bar{\eta}}}{\extd s} G_{k,\bar{\eta} \Phi}\,. \label{equationderivds}
\end{align}
Because of the causality condition
$
G_{k,\bar{\eta}\Phi}(\omega)=G_{k,\bar{\Phi}\eta}(-\omega)\,.
$
Note that in the Ward identities, we integrate over frequency and then the four contributions in \eqref{equationderivds} are indeed equal pairwise, and\footnote{We have explicitly shown here the components $\eta \bar{\Phi}$ and $\bar{\eta}\Phi$ for the regulator, the two components being equal to $R_k$.}
\begin{align}
\nonumber \frac{1}{2}\int \extd \omega\frac{\extd}{\extd s}\,G_{k,\bar{\Phi}\Phi}(\omega)=&-\int \extd \omega \,G_{k,\bar{\Phi}\eta}(\omega) \frac{\extd \Gamma_{k,\eta \bar{\Phi}}^{(2)}(\omega)}{\extd s} G_{k,\bar{\Phi} \Phi}(\omega)\\
&-i\int \extd \omega\, G_{k,\bar{\Phi}\eta}(\omega) \frac{\extd R_{k,\eta\bar{\Phi}}}{\extd s} G_{k,\bar{\Phi} \Phi}(\omega)\,.\label{equationderivkey}
\end{align}
Once again, it is suitable to use the truncation for the computation of the last term, without additional assumptions as those we introduce to define the vertex expansion. Including the sum over the momenta $q$, we recover what we call $\mathcal{A}_2$ in \eqref{defA2}.
\medskip

The right-hand side of \eqref{equationderivkey} can be easily computed from the same observation given in \cite{lahoche20242}: The relevant contribution of the derivative is essentially for momenta around $k$. However, this assumption is not sufficient, because the regulator does not depend on the frequency. Additional constraints are required and come from causality and time reversal symmetry. This symmetry imposes that $\Gamma^{(2)}_{k,\bar{\Phi}\Phi}=0$ (a condition which comes also from the invariance dynamics by adding linear forces, see remark \ref{conditiontrue}). This condition imposes that $G_{k,\bar{\Phi}\Phi}\propto G_{k,\bar{\eta}\Phi}(\omega)G_{k,\bar{\eta}\Phi}(-\omega)$. Furthermore, due to the causality $G_{k,\bar{\eta}\Phi}(\omega)$ has a pole in the negative half complex plane, $\omega=-iA_k$, such that $G_{k,\bar{\Phi}\Phi} \propto B_k/(\omega^2+A^2_k)$, and the integral over $\omega$ is easy to perform. Because ($A_k>0$):
\begin{equation}
\int \,\extd \omega \frac{B_k}{\omega^2+A^2_k}=\frac{\pi B_k}{A_k}\,,
\end{equation}
where coefficients $A_k$ and $B_k$ are assumed to depends on the generalized momenta $q$. Then:
\begin{equation}
\frac{1}{2}\int \extd \omega \int \rho(q) \extd q\,\frac{\extd}{\extd s}\, \frac{B_k}{\omega^2+A^2_k}=\frac{\pi}{2}\int \rho(q) \extd q\,\frac{\extd}{\extd s}\, \frac{B_k(q)}{A_k(q)}.
\end{equation}
At this step,   far from the scale $k$, the regulator is zero, and it makes sense to replace $A_k(q)/B_k(q)$ by $\tilde{Z} f_k(0,q)$:
\begin{align}
\frac{1}{2}\int \extd \omega \int \rho(q) \extd q\,\frac{\extd}{\extd s}\,G_{k,\bar{\Phi}\Phi}(0,q,\omega)=\frac{\pi}{2}\int \rho(q) \extd q \,\frac{\extd}{\extd s}\, \frac{1}{\tilde{Z} f_k(0,q)}\,.\label{equationPRIME}
\end{align}
The remaining integral may be computed easily  (see \cite{lahoche20242} for more detail). Finally, for $k$ small enough:
\begin{align}
\nonumber\frac{1}{2}\int \extd \omega \int &\rho(q) \extd q\frac{\extd}{\extd s}\,G_{k,\bar{\Phi}\Phi}(0,q,\omega)\approx \frac{\pi}{2\tilde{Z}}\Bigg[-\tilde{\nu} g(k,\tilde{u}_2)\\\nonumber
&+ \bigg(\frac{\sqrt{k}}{6 \pi  \left({\sigma }\right)^{3/2} \left(1+\tilde{u}_2\right)}+\frac{\sqrt{k}}{2 \sigma^{3/2}} \mathrm{F}_\pm(\tilde{u}_2)+\frac{\tilde{u}_2 k}{\sigma^2} \bigg) \\
&+\bigg(-\frac{\sqrt{k}}{3 \pi  \left({\sigma }\right)^{3/2} \left(1+\tilde{u}_2\right)^2}+\frac{\sqrt{k}}{\sigma^{3/2}} \mathrm{F}_\pm^\prime(\tilde{u}_2)+\frac{k}{\sigma^2}  \bigg)\dot{\tilde{u}}_2\Bigg]\,,\label{KEYSUM}
\end{align}
where  we  use equation \eqref{flowequ2OOF} at equilibrium, $\tilde{\nu}:= \frac{\extd}{\extd s} \ln \tilde{Z}\approx \nu$, and in the deep IR:
\begin{equation}
g(k,\tilde{u}_2)=\frac{\sqrt{k}}{3 \pi  \left({\sigma }\right)^{3/2} \left(1+\tilde{u}_2\right)}+\frac{1}{\sigma}+\frac{\sqrt{k}}{ \sigma^{3/2}} \mathrm{F}_\pm(\tilde{u}_2)+\frac{\tilde{u}_2 k}{\sigma^2} +\mathcal{O}(k^2)
\end{equation}

Now, using the Wetterich flow equation, we have \textit{in the symmetric phase}:
\begin{equation}
\dot{\Gamma}_{k,\eta \bar{\Phi}}^{(2)}(\vec{p}\,)= -2 \Gamma_{k,\eta \bar{\Phi} \Phi \bar{\Phi}}^{(1+\bar{2},1)}(\vec{p},\vec{p},\vec{0},\vec{0}\,)\, \N \underbrace{\int \extd \omega \int \rho(p) \extd p\, G_{k,\bar{\Phi}\Phi}(0,p,\omega) G_{k,\bar{\Phi}\eta}(0,p,\omega) i\dot{R}_k(0,p)}_{:=\mathcal{A}_2(k)}\,,
\end{equation}
where $\mathcal{A}_2(k)$ is given in the relation \eqref{A2result}), and we define $\mathcal{L}_2$ already introduced in the equilibrium dynamic on  \cite{lahoche20242}:
\begin{align}
\mathcal{A}_2(k)=\frac{1}{8 \tilde{Z}} \frac{1}{k}\frac{J_{\tilde{\nu}}}{(1+\tilde{u}_2)^2}=:\frac{\pi}{2}\, \mathcal{L}_2\,.\label{eqA2}
\end{align}
Then:
\begin{align}
\nonumber & 2\N\int \rho(q) \frac{\extd R_k}{\extd p_1}(0,q)\int \extd \omega G_{k,\bar{\Phi}\Phi}(0,q,\omega) G_{k,\bar{\eta}\Phi}(0,q,\omega)\gamma_{k,q000}^{(2+\bar{1},\bar{1})}\\\nonumber
&=- \frac{2}{\mathcal{A}_2}\,  \int \rho(q) \int \extd \omega \,G_{k,\bar{\Phi}\eta}(\omega) \frac{\extd \Gamma_{k,\eta \bar{\Phi}}^{(2)}(\omega)}{\extd s} G_{k,\bar{\Phi} \Phi}\\\nonumber
&= \frac{2}{\mathcal{A}_2} \int \rho(p) \int \extd \omega \Bigg[  \frac{1}{2}\int \extd \omega\frac{\extd}{\extd s}\,G_{k,\bar{\Phi}\Phi}(\omega)+i\int \extd \omega\, G_{k,\bar{\Phi}\eta}(\omega) \frac{\extd R_{k,\eta\bar{\Phi}}}{\extd s} G_{k,\bar{\Phi} \Phi}(\omega)\Bigg]\\\nonumber
&\approx \frac{2}{\mathcal{L}_2}\frac{1}{\tilde{Z}}\Bigg[-\tilde{\nu} g(k,\tilde{u}_2)+ \bigg(\frac{\sqrt{k}}{6 \pi  \left({\sigma }\right)^{3/2} \left(1+\tilde{u}_2\right)}+\frac{\sqrt{k}}{2 \sigma^{3/2}} \mathrm{F}_\pm(\tilde{u}_2)+\frac{\tilde{u}_2 k}{\sigma^2} \bigg) \\
&\quad +\bigg(-\frac{\sqrt{k}}{3 \pi  \left({\sigma }\right)^{3/2} \left(1+\tilde{u}_2\right)^2}+\frac{\sqrt{k}}{\sigma^{3/2}} \mathrm{F}_\pm^\prime(\tilde{u}_2)+\frac{k}{\sigma^2}  \bigg)\dot{\tilde{u}}_2\Bigg]+2\,,
\end{align}
and finally, Ward's identity reads:
\begin{align}
&\nonumber\frac{2}{\mathcal{L}_2}\frac{1}{\tilde{Z}}\Bigg[-\tilde{\nu} g(k,\tilde{u}_2)+ \bigg(\frac{\sqrt{k}}{6 \pi  \left({\sigma }\right)^{3/2} \left(1+\tilde{u}_2\right)}+\frac{\sqrt{k}}{2 \sigma^{3/2}} \mathrm{F}_\pm(\tilde{u}_2)+\frac{\tilde{u}_2 k}{\sigma^2} \bigg) \\\nonumber
&+\bigg(-\frac{\sqrt{k}}{3 \pi  \left({\sigma }\right)^{3/2} \left(1+\tilde{u}_2\right)^2}+\frac{\sqrt{k}}{\sigma^{3/2}} \mathrm{F}_\pm^\prime(\tilde{u}_2)+\frac{k}{\sigma^2}  \bigg)\dot{\tilde{u}}_2\Bigg]\\
&+2- \frac{u_4}{\tilde{Z}}\frac{\left(\frac{1}{\tilde{\sigma }}\right)^{3/2}}{6 k^2}\frac{1}{(1+\tilde{u}_2)^2}+(\tilde{Z}-1)=0\,.\label{eqWARD}
\end{align}
This result allows us to provide another proof that flow equations in this case match with the flow equation for the equilibrium theory (note that we set $T=1$ in this paper, whereas we fixed $T=2$ in reference \cite{lahoche20242}). Remark that, despite  the flow equations  are the same, the interpretations are different. To be more precise in the deep IR, the anomalous dimension $\tilde{\nu}$ has to vanishes (i.e. $\tilde{Z}=1$), and using \eqref{flowequ2OOF}, we are able to express $\bar{u}_4$ in function of $\bar{u}_2$ only, and we denote as $\bar{u}_4^*(\bar{u}_2)$ this solution. In turn, it allows closing the hierarchy from \eqref{flowequ2OOF}, which becomes a differential equation for $\bar{u}_2$ only:

\begin{equation}
\boxed{\dot{\tilde{u}}_2=-(1+\nu)\tilde{u}_2-\frac{7\bar{u}_4^*(\tilde{u}_2)}{24 \sigma^{3/2}}\, \frac{1}{(1+\tilde{u}_2)^2}\,.}\label{flowequ2OOFOS}
\end{equation}
where $\bar{u}_4:=\sqrt{k^3}\, \tilde{u}_4$.

\section{The  out of equilibrium dynamics}\label{sec3}
In this section, we investigate a study beyond the equilibrium dynamics, and we consider a parametrization of the theory space allowing violations of the fluctuation-dissipation theorem. As in the previous section, we  use of the Ward identities to close the hierarchy, and we express all the couplings in the function of $\bar{u}_2$, in the deep IR and, in the symmetric phase.

\subsection{Explicit breaking of the fluctuation-dissipation theorem}
\begin{figure}
\begin{center}
\includegraphics[scale=1]{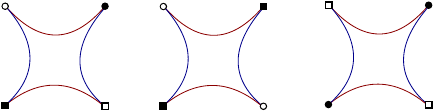}
\end{center}
\caption{The three quartic interactions involving two response fields.}\label{figtworesponseexample}
\end{figure}

 The fluctuation-dissipation theorem \eqref{fluctuationdissipationTH} is a direct consequence of the form of the matrix \eqref{matrixGamma2}, namely:
 \begin{equation}
\Gamma^{(2)}_k(\vec{p},\omega) = \begin{pmatrix}
A& B(\vec{p},\omega) \\
B(\vec{p},-\omega)& 0
\end{pmatrix}
\,.
 \end{equation}
The time-reversal symmetry implies that: if $(\Gamma^{(2)}_k)_{\bar{\eta} \eta}=A$, $B(\vec{p},\omega)=- A \omega + i f_k(\vec{p})$ i.e the wave function renormalization for the response field equals the wave function renormalization for $\bar{\eta} \dot{\Phi}$. From this observation, we expect that a simple way to break the time-reversal symmetry (and then the fluctuation-dissipation theorem) is to choose a truncation where these two coefficients are not equal, for instance in the case where  $A\neq 1$ and $B(\vec{p},\omega)=-\omega + i f_k(\vec{p}\,)$. To this end, we have to enrich the truncation such that the flow of $A$ becomes non-trivial, allowing the interactions having two response fields instead of one, as Figure \ref{figtworesponseexample} shows. The two interactions on the right do not contribute to the vertex $\bar{\eta}\eta$ since black and white nodes have to be contracted pairwise. The first one on the left, however, contributes to the vertex $\bar{\eta}\eta$, but also includes a correction for the $2$-points vertex $\bar{\eta}\Phi$, ensuring that the flow for mass and coupling $(\Gamma^{(2)}_k)_{\bar{\eta} \eta}$ are not independents. We can then show that the flow equation for $u_2$ is essentially unchanged due to the causality. Then, we consider a truncation of the form:
\begin{align}
\nonumber \Gamma_{k,\text{int}}[\eta,\bar{\eta},\Phi,\bar{\Phi}]:=& \sum_p i\frac{(1+\delta_{1p})u_{2p}}{p!(p-1)!}\, \left(\overbrace{\vcenter{\hbox{\includegraphics[scale=0.65]{figinteraction1.pdf}}}}^{\text{order} \,2p}\,+\,\overbrace{\vcenter{\hbox{\includegraphics[scale=0.65]{figinteraction2.pdf}}}}^{\text{order}\, 2p}\right)\\
& +\sum_{p>1}\, \frac{v_{2p}}{(p-1)!^2} \, \left(\overbrace{\vcenter{\hbox{\includegraphics[scale=0.65]{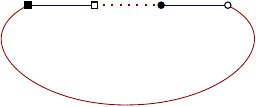}}}}^{\text{order} \,2p}\,\right)\,,\label{expansionGammaBIS}
\end{align}
with the kinetic kernel:
\begin{equation}
\Gamma_k^{(2)}(p_1,p_2,\omega)=\begin{pmatrix}
1+\Delta + \Delta^\prime(p_1+p_2) & -\omega+iZ(k)(p_1+p_2+2k\tilde{u}_2) \\
\omega+iZ(k)(p_1+p_2+2k\tilde{u}_2)& 0
\end{pmatrix}
\,.\label{matrixGamma2}
\end{equation}
The quantity $\Delta$ measures the violation of the time-reversal symmetry. The flow equation for $u_2$ and $\Delta$ can be deduced as it is the case  for $\bar{u}_2$ in section \ref{secfloweq}. Graphically, the flow equations read:
\begin{equation}
\dot{u}_2\,=\,-\,\vcenter{\hbox{\includegraphics[scale=0.8]{loop1}}}\,-\,\vcenter{\hbox{\includegraphics[scale=0.8]{loop2}}}\,-\,\vcenter{\hbox{\includegraphics[scale=0.8]{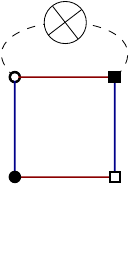}}}\,-\,\vcenter{\hbox{\includegraphics[scale=0.8]{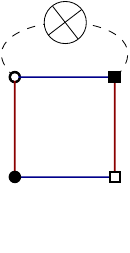}}}\,,\label{floweq2ptsOOEBIS}
\end{equation}
\begin{equation}
\dot{\Delta}\,=\,-\,\vcenter{\hbox{\includegraphics[scale=0.8]{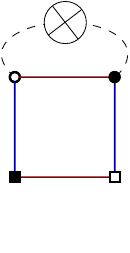}}}\,-\,\vcenter{\hbox{\includegraphics[scale=0.8]{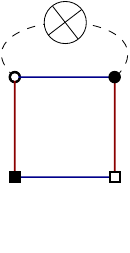}}}\,.
\end{equation}
Let's focus first on the flow equation for $\bar{u}_2$. The two first terms have been computed previously, and the two last ones involve the integral:
\begin{equation}
\mathcal{B}_2:=\int \, \extd \omega \, \int \, \rho(p) \extd p \,\frac{\dot{R}_k(p,0)}{(\omega+i Z f_k(p,0))^2}\,=0\,.\label{defB2}
\end{equation}
that is once again a consequence of causality $G_{k,\bar{\eta}\Phi}(t,t^\prime) \propto \theta(t-t^\prime)$, and the loop involves the product $\theta(t-t^\prime)\theta(t^\prime-t)$, which vanishes. Then the flow equation for $\tilde{u}_2$ remains essentially the same as \eqref{flowequ2OOF}, except for an additional factor $(1+\Delta)$ arising from the modification of the component $\bar{\eta}\eta$ of $\Gamma_k^{(2)}$:
\begin{equation}
\boxed{\dot{\tilde{u}}_2=-(1+\nu)\tilde{u}_2-\frac{\tilde{u}_4}{8}\, \frac{(1+\Delta)J_{{\nu}}(k)+\bar{\Delta}^\prime K_{{\nu}}(k)}{(1+\tilde{u}_2)^2}\,,}\label{flowequ2OOF}
\end{equation}
where:
\begin{align}
K_\nu&= \frac{1}{2\tilde{\sigma}^2} \left[\sqrt{4\tilde{\sigma}-1}+\int_0^1 \extd q q\sqrt{q(4\tilde{\sigma}-q)}\left({\nu}(2-q)+2\right)\right]\,,\label{defK}
\end{align}
and:
\begin{equation}
\bar{\Delta}^\prime:=k \Delta^\prime\,.
\end{equation}
The flow equation for\footnote{The observable $\Delta$ is dimensionless by construction.} $\bar{\Delta}\equiv \Delta$ involves the integral $\mathcal{A}_2$ (Eq. \eqref{A2result}), multiplied by the factor $1+\Delta$, and we get in the deep IR:
\begin{align}
\boxed{\dot{\Delta}= -\frac{\tilde{v}_4}{4}\, \frac{(1+\Delta)J_{{\nu}}(k)+\bar{\Delta}^\prime K_{{\nu}}(k)}{(1+\tilde{u}_2)^2}\,.}\label{dotDelta}
\end{align}
where we defined $\tilde{v}_4:= v_4 k^{-1} Z^{-1}(k)$. Finally, if we neglect the momenta dependency of the effective vertices, the flow equation for $\Delta^\prime$ can be computed following the same method as we used in \cite{lahoche20241} to compute the anomalous dimension. We have:
\begin{align}
\dot{\Delta}^\prime &=  - i \frac{v_4}{2} \frac{\extd}{\extd q}\int \, \extd \omega \, \int \, \rho(p) \extd p \frac{\dot{R}_k(p,q)}{\omega^2+Z^2f^2_k(p,0)}\frac{1+\Delta+\Delta^\prime(p+q)}{\omega+i Z f_k(p,0)} \,\Big\vert_{q=0}\\
&=- \frac{\pi v_4}{16} \frac{\extd}{\extd q} \, \int \, \rho(p) \extd p \frac{\dot{R}_k(p,q)}{(k+u_2)^2}(1+\Delta+\Delta^\prime(p+q))\,\Big\vert_{q=0}.
\end{align}
To compute the second line of the above expression, we used explicitly the fact that the anomalous dimension must vanish in the deep IR, as we will see below.  Then using the relation  $
 \frac{\extd}{\extd q}\dot{R}_k(p,q)\,\Big\vert_{q=0}=-k \delta(k-p)\,,
$
in the deep IR:
\begin{equation}
\boxed{\dot{\bar{\Delta}}^\prime= \bar{\Delta}^\prime+ \frac{\tilde{v}_4}{16\tilde{\sigma}^{3/2}}\frac{1}{(1+\tilde{u}_2)^2}\left(1+\Delta- \frac{4}{3}\,\bar{\Delta}^\prime \right)\,.}\label{flotdeltaprime}
\end{equation}

\subsection{Ward identities and deep IR flow}
In this section 
 we will  use the Ward identity  to close the hierarchy, as we have done in the previous section for equilibrium dynamics in the deep IR. First, let us investigate how the Ward identity constraint \eqref{eqWARD} changes due to the presence of $\Delta$ and $v_4$. First, the integrals involving $G_{k,\bar{\Phi}\Phi}$ have to be multiplied by $1+\Delta$. Second, one may expect that many components have to be added for expansions like \eqref{equationderivkey} and \eqref{decompositionG1}-\eqref{decompositionG2}. However, a moment of reflection shows that equation \eqref{decompositionG1} does not change as we discard terms like $\Gamma^{(1+1;\bar{2})}_{k}$ from the truncation. Equation \eqref{decompositionG2}  reads explicitly:
\begin{align}
\frac{\partial^2G_{k,\bar{\Phi}\eta}(\vec{p}\,^\prime,{\omega};\vec{p},{\omega})}{\partial \Phi_{\bm q}({\omega}_1)\partial \bar{\eta}_{\bm{\bar{q}}}({\bar{\omega}}_1)}=&- \sum_{p_1,p_1^\prime} G_{k,\bar{\Phi}\Phi}(p^\prime,p_1^\prime) \Gamma_{k,\Phi\bar{\Phi}\Phi\bar{\eta}}^{(4)}(q,\bar{q},p_1^\prime,p_1) G_{k,\bar{\Phi} \eta}(p_1,p)\\
&- \sum_{p_1,p_1^\prime} G_{k,\bar{\Phi}\eta}(p^\prime,p_1^\prime) \Gamma_{k,\eta\bar{\Phi}\Phi\bar{\eta}}^{(4)}(q,\bar{q},p_1^\prime,p_1) G_{k,\bar{\Phi} \eta}(p_1,p)\,,\label{decompositionG2prime}
\end{align}
but the last term involves a frequency integral like $\mathcal{B}_2$ and vanishes due to the causality properties. Finally, \eqref{equationderivkey} becomes:
\begin{align}
\nonumber \frac{\extd}{\extd s}\,G_{k,\bar{\Phi}\Phi}=&-G_{k,\bar{\Phi}\eta} \frac{\extd \Gamma_{k,\eta \bar{\eta}}^{(2)}}{\extd s} G_{k,\bar{\eta} \Phi}\\\nonumber
&-G_{k,\bar{\Phi}\eta} \frac{\extd \Gamma_{k,\eta \bar{\Phi}}^{(2)}}{\extd s} G_{k,\bar{\Phi} \Phi}-G_{k,\bar{\Phi}\eta} \frac{\extd R_{k,\eta\bar{\Phi}}}{\extd s} G_{k,\bar{\Phi} \Phi}\\
& - G_{k,\bar{\Phi}\Phi} \frac{\extd \Gamma_{k,\Phi \bar{\eta}}^{(2)}}{\extd s} G_{k,\bar{\eta} \Phi}-G_{k,\bar{\Phi}\Phi} \frac{\extd R_{k,\Phi\bar{\eta}}}{\extd s} G_{k,\bar{\eta} \Phi}\,. \label{equationderivdsBIS}
\end{align}
The added contribution in the first line does not vanish, because $G_{k,\bar{\Phi}\eta} (\omega)=G_{k,\bar{\eta} \Phi} (-\omega)$. But let us define:
\begin{equation}
\bar{G}_{k,\bar{\Phi}\Phi}:= \frac{1}{\Gamma_{k,\bar{\eta}\eta}} {G}_{k,\bar{\Phi}\Phi}\,,
\end{equation}
Such that:
\begin{align}
\nonumber \Gamma_{k,\bar{\eta}\eta} \frac{\extd}{\extd s}\,\bar{G}_{k,\bar{\Phi}\Phi}=&-G_{k,\bar{\Phi}\eta} \frac{\extd \Gamma_{k,\eta \bar{\Phi}}^{(2)}}{\extd s} G_{k,\bar{\Phi} \Phi}-G_{k,\bar{\Phi}\eta} \frac{\extd R_{k,\eta\bar{\Phi}}}{\extd s} G_{k,\bar{\Phi} \Phi}\\
& - G_{k,\bar{\Phi}\Phi} \frac{\extd \Gamma_{k,\Phi \bar{\eta}}^{(2)}}{\extd s} G_{k,\bar{\eta} \Phi}-G_{k,\bar{\Phi}\Phi} \frac{\extd R_{k,\Phi\bar{\eta}}}{\extd s} G_{k,\bar{\eta} \Phi}\,. 
\end{align}
But on-shell, i.e. along the portion of the theory space spanned by the truncation, $\bar{G}_{k,\bar{\Phi}\Phi}$ is nothing but the equilibrium $\bar{\Phi}\Phi$ component of the two point function. However, we cannot use  the same argument as before, in the equilibrium dynamics regime, because of the factor $ \Gamma_{k,\bar{\eta}\eta} $. Indeed, the approximation requires that $ \Gamma_{k,\bar{\eta}\eta} =1+\Delta$ only in the windows of momenta around $k$ allowed by the regularization, but not outside. Finally, we cannot in principle replace $ \Gamma_{k,\bar{\eta}\eta} $ by $1+\Delta$ in the computation of the sum. A solution consists of deriving with respect to $s$ the identity and using the fact that sums and integrals are convergent to exchange the order of derivatives and sums. In that way, we recover again the derivative of a $k$ dependent sum, and  we expect that the relevant contribution comes from momenta around $k$, such that it makes sense \textit{a priori} to replace quantities by their expressions from the truncation. 
\medskip

\begin{figure}
\begin{center}
\includegraphics[scale=0.8]{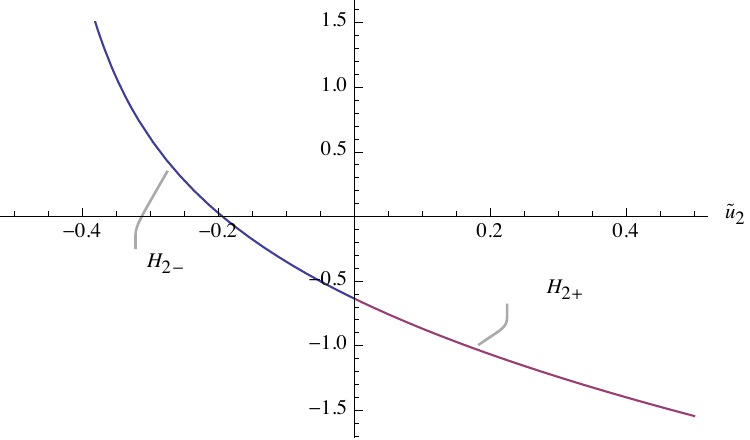}
\end{center}
\caption{Behavior of the function $H_{2\pm}$ near $\tilde{u}_2=0$.}\label{FIGH}
\end{figure}

Our proposed calculation method, however, has the disadvantage of complicating the equations and poses the question of the integration constant. An equivalent method consists of noticing that in the calculation of a term like: 
\begin{equation}
\int \extd \omega\sum_p \, \Gamma_{k,\bar{\eta}\eta}(p,0) \frac{\extd}{\extd s}\,\bar{G}_{k,\bar{\Phi}\Phi}(p,0)\,,
\end{equation}
the relevant contribution comes from the momenta around $k$, because far from this momentum scale, the influence of the regulator is negligible, and the derivative is almost zero. Then, in the deep IR, it makes sense to replace $\Gamma_{k,\bar{\eta}\eta}(p,0)$ by $1+\Delta+\Delta^\prime p$\,. Explicitly:
\begin{align}
\frac{1}{2}\int \extd \omega& \int \rho(q) \extd q\,\Gamma_{k,\bar{\eta}\eta}(p,0)\,\frac{\extd}{\extd s}\,\bar{G}_{k,\bar{\Phi}\Phi}(0,q,\omega)\\
&=\frac{\pi}{2}\int \rho(q) \extd q \,(1+\Delta+\Delta^\prime q)\,\frac{\extd}{\extd s}\, \frac{1}{ \tilde{Z} f_k(0,q)}\\
&=-\frac{\pi}{2} \,\int \rho(q) \extd q \,(1+\Delta+\Delta^\prime q)\, \frac{\tilde{Z} \tilde{\nu} q+2\dot{u}_2+\dot{R}_k}{ (\tilde{Z} f_k(0,q))^2}\,,\label{equationSECOND}
\end{align}
The term proportional to $1+\Delta$ is exactly  computed before (see equation \eqref{KEYSUM}). Furthermore:
\begin{align}
\frac{\pi \Delta^\prime}{2} \,\int \rho(q) \extd q \,q\, \frac{\tilde{Z} \nu q+2\dot{u}_2+\dot{R}_k}{ (\tilde{Z} f_k(0,q))^2}=\frac{\pi \Delta^\prime}{8 \tilde{Z}} \left[\frac{K_\nu}{\pi (1+\tilde{u}_2)^2}+\tilde{\nu} h_1(k,\tilde{u}_2)+\frac{2\dot{u}_2}{k \tilde{Z}} k h_2(k,\tilde{u}_2)\right] \,,
\end{align}
where:
\begin{align}
\nonumber h_1(k,\tilde{u}_2)=\frac{\left(\frac{k}{\sigma }\right)^{3/2}}{14 \pi  (1+\tilde{u}_2)^2}+1-\frac{4 k \tilde{u}_2}{\sigma }+\mathcal{O}\left(k^{3/2}\right)\,,
\end{align}
and:
\begin{align}
 h_2(k,\tilde{u}_2)=\frac{\sqrt{k}}{10 \pi  \sigma ^{3/2} (1+\tilde{u}_2)^2}+\sqrt{k}H_{2\pm}(\tilde{u}_2)-\frac{k}{2 \sigma ^2}+\mathcal{O}\left(k^{3/2}\right)\,,
\end{align}
with $H_{2\pm}(\tilde{u}_2)$ given by
\begin{equation}
H_{2\pm}(\tilde{u}_2) = \left\{
    \begin{array}{ll}
        \frac{-3 \sqrt{2} \sqrt{\sigma  \tilde{u}_2}-\frac{4 \sqrt{\sigma } \left(3 \tilde{u}_2+1\right)}{2 \pi  \tilde{u}_2+\pi }+\frac{6 \sqrt{2} \sqrt{\sigma  \tilde{u}_2} \tan ^{-1}\left(\frac{1}{\sqrt{2} \sqrt{\tilde{u}_2}}\right)}{\pi }}{2 \sigma ^2} & \mbox{if} \quad \tilde{u}_2>0 \\
     \frac{\frac{3 i \sqrt{-\tilde{u}_2}}{\sqrt{2}}-\frac{2 \left(3 \tilde{u}_2+1\right)}{2 \pi  \tilde{u}_2+\pi }+\frac{3 \sqrt{2} \sqrt{\tilde{u}_2} \tan ^{-1}\left(\frac{1}{\sqrt{2} \sqrt{\tilde{u}_2}}\right)}{\pi }}{\sigma ^{3/2}} & \mbox{if}\quad  \tilde{u}_2<0\,.
    \end{array}
\right.
\end{equation}
Figure \ref{FIGH} shows the behavior of functions $H_{2\pm}$. 
Finally, in the deep IR as for equation \eqref{equationtermcool}, we get
\begin{align}
\nonumber\int \rho(q)& \frac{\extd R_k}{\extd p_1}(0,q)\int \extd \omega G_{k,\bar{\Phi}\Phi}(0,q,\omega) G_{k,\bar{\eta}\Phi}(0,q,\omega)\gamma_{k,q000}^{(2+\bar{1},\bar{1})}\\\nonumber
&=-\frac{i \pi}{2 \tilde{Z}}\gamma_{k,0000}^{(2+\bar{1},\bar{1})} \times \int \rho(q)\,\frac{\extd R_k}{\extd p_1}(0,q) \frac{1+\Delta+\Delta^\prime \, q}{(f_k(0,q))^2}\\
&= -\frac{u_4}{\N\tilde{Z}}\frac{\left(\frac{1}{\tilde{\sigma }}\right)^{3/2}}{12 k^2}\frac{1+\Delta+\frac{3}{5}\bar{\Delta}^\prime}{(1+\tilde{u}_2)^2}\,.\label{equationtermcoolBIS}
\end{align}
Furthermore, the definition of $\mathcal{A}_2$ is also modified as:
\begin{align}
\mathcal{A}_2(k)=\frac{1}{8 \tilde{Z}} \frac{1}{k}\frac{(1+\Delta)J_{{\nu}}(k)+\bar{\Delta}^\prime K_{{\nu}}(k)}{(1+\tilde{u}_2)^2}=:\frac{\pi}{2}\, \mathcal{L}_2\,.\label{eqA2BIS}
\end{align}
Now defining the functions $W_{\pm}^{(0)}(k,\tilde{u}_2,\tilde{\nu})$ and $W_{\pm}^{(1)}(k,\tilde{u}_2,\tilde{\nu})$ as:
\begin{align}
\nonumber W_{\pm}^{(0)}(k,\tilde{u}_2,\tilde{\nu})&:=\frac{2}{\mathcal{L}_2}\frac{1}{\tilde{Z}}\Bigg[-\tilde{\nu} g(k,\tilde{u}_2)+ \bigg(\frac{\sqrt{k}}{6 \pi  {\sigma}^{3/2} \left(1+\tilde{u}_2\right)}+\frac{\sqrt{k}}{2 \sigma^{3/2}} \mathrm{F}_\pm(\tilde{u}_2)+\frac{\tilde{u}_2 k}{\sigma^2} \bigg) \\
&+\bigg(-\frac{\sqrt{k}}{3 \pi{\sigma}^{3/2} \left(1+\tilde{u}_2\right)^2}+\frac{\sqrt{k}}{\sigma^{3/2}} \mathrm{F}_\pm^\prime(\tilde{u}_2)+\frac{k}{\sigma^2}  \bigg)\dot{\tilde{u}}_2\Bigg]+2\,,
\end{align}
and:
\begin{equation}
W_{\pm}^{(1)}(k,\tilde{u}_2,\tilde{\nu}):=\frac{-1}{2\mathcal{L}_2\tilde{Z}} \left[\frac{K_\nu}{\pi (1+\tilde{u}_2)^2}+\tilde{\nu} h_1(k,\tilde{u}_2)+\frac{2\dot{u}_2}{k \tilde{Z}} h_2(k,\tilde{u}_2)\right]\,,
\end{equation}
the Ward identity reads finally as:
\begin{align}
\boxed{(1+\Delta)W_{\pm}^{(0)}(k,\tilde{u}_2,\tilde{\nu}) +\Delta^{\prime}W_{\pm}^{(1)}(k,\tilde{u}_2,\tilde{\nu})-\frac{u_4}{\tilde{Z}}\frac{\left(\frac{1}{\tilde{\sigma }}\right)^{3/2}}{6 k^2}\frac{1+\Delta+\frac{3}{5}\bar{\Delta}^\prime}{(1+\tilde{u}_2)^2}+\tilde{Z}-1=0\,.}\label{eqWARDBIS}
\end{align}
In the equilibrium theory, we investigated the leading order contribution in front of the anomalous dimension, of order $1/\sqrt{k}$ and the cancellation of this contribution provided the constraint $\tilde{\nu}=0$ in the deep IR. Here, the leading order contribution comes from the term in front of $\bar{\Delta}^\prime$, but the divergence is  of the form $1/k$ for $k$ goes to infinity, and the cancellation of this contribution imposes in the deep IR:
\begin{equation}
\boxed{\tilde{\nu}\bar{\Delta}^\prime=0\,.}
\end{equation}
In the same way, the cancellation of the term of order $1/\sqrt{k}$  requires that $\tilde{\nu}=0$ (assuming $\Delta \neq -1$), and the Ward identity simplifies as :
\begin{align}
\boxed{(1+\Delta)W_{\pm}^{(0)}(k,\tilde{u}_2,0) +\Delta^{\prime}W_{\pm}^{(1)}(k,\tilde{u}_2,0)-\frac{\tilde{u}_4}{6 }\left(\frac{1}{\tilde{\sigma }}\right)^{3/2}\frac{1+\Delta+\frac{3}{5}\bar{\Delta}^\prime}{(1+\tilde{u}_2)^2}=0\,.}\label{eqWARDBIS}
\end{align}
This relation is nothing but the equilibrium condition,  and the solution $\tilde{u}_4^*(\tilde{u}_2)$ remains the same as we deduce from \eqref{eqWARD}  with the definition of $\mathcal{L}_2$ which becomes different.
\medskip
Now, let us investigate the relation coming from Ward identity \eqref{WardIDoutofEq2}, taking derivative with respect to $\delta^2/\delta \eta_{\vec{q}\,}(\Omega)\delta \bar{\eta}_{\vec{\bar{q}}\,}(\bar{\Omega})$:
\begin{align}
\nonumber \int \extd\omega & \sum_{p_2} \Bigg[ i \delta E_k(p_1,p_1^\prime,p_2) \Bigg( \frac{\delta^2 G^{(\bar{1};1)}_{k,\bar{\eta} \Phi}(\vec{p}\,^\prime,\omega; \vec{p},\omega)}{\delta \eta_{\vec{q}}(\Omega)\delta \bar{\eta}_{\vec{\bar{q}}}(\bar{\Omega})}
+\frac{\delta^2 G^{({1};\bar{1})}_{k,\bar{\Phi} \eta}(\vec{p},\omega; \vec{p}\,^\prime,\omega)}{\delta \eta_{\vec{q}}(\Omega)\delta \bar{\eta}_{\vec{\bar{q}}}(\bar{\Omega})}\Bigg) \\
&+\Bigg(\frac{\delta\bar{\tilde{L}}_{\vec{p}\,}(\omega)}{\delta \bar{\eta}_{\vec{\bar{q}}}(\bar{\Omega})}\delta_{\vec{p}\,^\prime \vec{q}\,}\delta(\omega-\Omega)-  \frac{\delta{\tilde{L}}_{\vec{p}\,^\prime}(\omega)}{\delta \eta_{\vec{q}}(\Omega)} \delta_{\vec{p}\, \vec{\bar{q}}\,}\delta(\omega-\bar{\Omega})\Bigg)\Bigg]=0\,.\label{Wardclosing2BIS}
\end{align}
The last contributions can be easily computed:
\begin{equation}
\frac{\delta\bar{\tilde{L}}_{\vec{p}\,}(\omega)}{\delta \bar{\eta}_{\vec{\bar{q}}\,}(\bar{\Omega})}=\Gamma_{k,\bar{\eta}\eta}^{(2)}(\omega,\vec{p}\,) \delta_{\vec{p}\,\vec{\bar{q}}\,}\delta(\omega-\Omega)\,.
\end{equation}
Then, in the limit $\vec{p}\to \vec{p}\,^\prime\to 0$, the second term is equal to $\Delta^\prime=0$. The Ward identity then simplifies as:
\begin{equation}
i\int \extd\omega \sum_{p_2} \frac{\extd}{\extd p_1} E_k(p_1,0,p_2)\Big\vert_{p_1=0}\, \Bigg( \frac{\delta^2 G_{k,\bar{\eta} \Phi}(\vec{p}\,^\prime,\omega; \vec{p},\omega)}{\delta \eta_{\vec{q}}(\Omega)\delta \bar{\eta}_{\vec{\bar{q}}}(\bar{\Omega})}
+\frac{\delta^2 G_{k,\bar{\Phi} \eta}(\vec{p},\omega; \vec{p}\,^\prime,\omega)}{\delta \eta_{\vec{q}}(\Omega)\delta \bar{\eta}_{\vec{\bar{q}}}(\bar{\Omega})}\Bigg) =\Delta^\prime \,.
\end{equation}
The two functional derivatives can be easily computed. Schematically, we have:
\begin{equation}
\frac{\delta^2 G_{k,\bar{\eta} \Phi}}{\delta \eta \delta \bar{\eta}}=-G_{k,\bar{\Phi}\Phi} \Gamma_{k}^{(1+\bar{1};1+\bar{1})} G_{k,\bar{\eta}\Phi}\propto  G_{k,\bar{\Phi}\Phi} \dot{\Gamma}_{k,\bar{\eta}\eta}^{(2)} G_{k,\bar{\eta}\Phi}
\end{equation}
and:
\begin{equation}
\frac{\delta^2 G_{k,\bar{\Phi} \eta}}{\delta \eta \delta \bar{\eta}}=-G_{k,\bar{\Phi}\Phi} \Gamma_{k}^{(1+\bar{1};1+\bar{1})} G_{k,\bar{\Phi}\eta}\propto  G_{k,\bar{\Phi}\Phi} \dot{\Gamma}_{k,\bar{\eta}\eta}^{(2)} G_{k,\bar{\Phi}\eta}\,,
\end{equation}
and because the function $G_{k,\bar{\Phi}\Phi}(\omega)$ is expected to be symmetric, 
\begin{equation}
\frac{\delta^2 G_{k,\bar{\eta} \Phi}}{\delta \eta \delta \bar{\eta}}+\frac{\delta^2 G_{k,\bar{\Phi} \eta}}{\delta \eta \delta \bar{\eta}}\propto 2 G_{k,\bar{\Phi}\Phi} \dot{\Gamma}_{k,\bar{\eta}\eta}^{(2)} G_{k,\bar{\eta}\Phi}\,.
\end{equation}
Furthermore, as previously, the term involving the derivative of the regulator can be computed using the truncation, and we get:

\begin{align}
\N i\int \rho(q) \int \extd \omega G_{k,\bar{\Phi}\Phi}(0,q,\omega) G_{k,\bar{\eta}\Phi}(0,q,\omega)\,\dot{\Gamma}_{k,\bar{\eta}\eta}^{(2)}(0,q) - \frac{v_4}{\tilde{Z}}\frac{\left(\frac{1}{\tilde{\sigma }}\right)^{3/2}}{12 k^2}\frac{1+\Delta}{(1+\tilde{u}_2)^2}=\Delta^{\prime}\,.\label{FMWI0MOUT2}
\end{align}
The remaining integral can be also computed from the principle we used to compute the previous Ward identities. We have:
\begin{equation}
\int \rho(q) \int \extd \omega G_{k,\bar{\Phi}\Phi}(0,q,\omega) G_{k,\bar{\eta}\Phi}(0,q,\omega)=-\frac{i \pi}{2 \tilde{Z}^2} \times \int \rho(q)\, \frac{1+\Delta}{(f_k(0,q))^2}\,,
\end{equation}
and the computation of the integral leads to:
\begin{equation}
\boxed{\left(\sqrt{k}\,\frac{\pi H_{1\pm}(\tilde{u}_2)}{\mathcal{A}_2(k)}\dot{\Delta}-\frac{2}{3}\left(\frac{1}{\tilde{\sigma }}\right)^{3/2}\frac{\tilde{v}_4}{(1+\tilde{u}_2)^2}\right)(1+\Delta)=\bar{\Delta}^\prime\,,}
\end{equation}
where:
\begin{equation}
H_{1\pm}(\tilde{u}_2)=\left\{
    \begin{array}{ll}
       \frac{\frac{3 \sqrt{2\sigma} }{\sqrt{\tilde{u}_2}}+2 \sqrt{\sigma } \left(\frac{6}{2 \pi  \tilde{u}_2+\pi }+\frac{1}{\pi  \left(\tilde{u}_2+1\right){}^2}\right)-\frac{6 \sqrt{2} \tan ^{-1}\left(\frac{1}{\sqrt{2} \sqrt{\tilde{u}_2}}\right)}{\pi  \sqrt{\sigma  \tilde{u}_2}}}{12 \sigma ^2} & \mbox{if} \quad \tilde{u}_2>0 \\
    \frac{\frac{3 i \sqrt{2}}{\sqrt{-\tilde{u}_2}}+\frac{2}{\pi  \left(\tilde{u}_2+1\right){}^2}+\frac{12}{2 \pi  \tilde{u}_2+\pi }-\frac{6 \sqrt{2} \tan ^{-1}\left(\frac{1}{\sqrt{2} \sqrt{\tilde{u}_2}}\right)}{\pi  \sqrt{\tilde{u}_2}}}{12 \sigma ^{3/2}} & \mbox{if}\quad  \tilde{u}_2<0\,,
    \end{array}
\right.
\end{equation}
and in the deep IR:
\begin{equation}
\frac{\mathcal{A}_2(k)}{\sqrt{k}}=\frac{7}{24 \sigma^{3/2}(1+\tilde{u}_2)^2}\left(1+\Delta+\frac{27}{35}\bar{\Delta}^\prime\right)\,.
\end{equation}
Using \eqref{dotDelta}, this equation provides us a solution for $\bar{\Delta}^\prime=U(\tilde{v}_4,\tilde{u}_2,\Delta)=:(1+\Delta)T(\tilde{v}_4,\tilde{u}_2)$, with:
\begin{equation}
T(\tilde{v}_4,\tilde{u}_2):=-2\left(\sigma^{3/2} \pi H_{1\pm}(\tilde{u}_2)+ \frac{1}{3}\frac{1}{(1+\tilde{u}_2)^2} \right)\frac{\tilde{v}_4}{\tilde{\sigma}^{3/2}}\,.
\end{equation}
It is suitable to redefine the quartic couplings as:
\begin{equation}
\bar{u}_4:=\sqrt{k^3}\, \tilde{u}_4\,,\qquad \bar{v}_4:=\sqrt{k^3}\, \tilde{v}_4\,.
\end{equation}
Differentiating with respect to $s$, and using \eqref{flotdeltaprime}, we get:
\begin{align}
\dot{\bar{v}}_4:=\frac{(1+\Delta)\Bigg(T(\bar{v}_4,\tilde{u}_2)+ \frac{\bar{v}_4}{16{\sigma}^{3/2}}\frac{1-\frac{4}{3}\,T(\bar{v}_4,\tilde{u}_2)}{(1+\tilde{u}_2)^2}\Bigg)- T(\bar{v}_4,\tilde{u}_2 )\dot{\Delta}-\partial_{\tilde{u}_2} U(\bar{v}_4,\tilde{u}_2,\Delta) \dot{\tilde{u}}_2}{ \partial_{\bar{v}_4} U(\bar{v}_4,\tilde{u}_2,\Delta) }\,,\label{equationbetaV4}
\end{align}
which define the $\beta$-function $\dot{\bar{v}}_4=:\beta_{\bar{v}_4}(\bar{v}_4,\tilde{u}_2,\Delta)$. In the same way, from \eqref{eqWARDBIS}, we deduce $\bar{u}_4=V(\tilde{u}_2,\bar{v}_4,\Delta)$, 
\begin{equation}
V(\tilde{u}_2,\bar{v}_4,\Delta)=:\frac{2 X(\tilde{u}_2,\bar{v}_4,\Delta)}{Y(\tilde{u}_2,\bar{v}_4,\Delta)}\,,
\end{equation}
with:
\begin{align}
\nonumber &X(\tilde{u}_2,\bar{v}_4,\Delta)=\Bigg[ \bigg(\frac{1}{6 \pi  {\sigma}^{3/2} \left(1+\tilde{u}_2\right)}+\frac{1}{2 \sigma^{3/2}} \mathrm{F}_\pm(\tilde{u}_2) \bigg) \\\nonumber
&-\bigg(-\frac{1}{3 \pi{\sigma}^{3/2} \left(1+\tilde{u}_2\right)^2}+\frac{1}{\sigma^{3/2}} \mathrm{F}_\pm^\prime(\tilde{u}_2) \bigg){\tilde{u}}_2\Bigg]+\frac{7(1+\Delta)}{12\sigma^{3/2}}\, \frac{1+\frac{27}{35}T(\bar{v}_4,\tilde{u}_2) }{(1+\tilde{u}_2)^2}\\
&-\frac{T(\bar{v}_4,\tilde{u}_2)}{5\pi \sigma^{3/2}(1+\tilde{u}_2)^2}\,,
\end{align}
and
\begin{align}
\nonumber Y(\tilde{u}_2,\bar{v}_4,\Delta)&:=-\frac{7(1+\Delta)}{12\sigma^{3/2}}\, \frac{1+\frac{27}{35}T(\bar{v}_4,\tilde{u}_2) }{(1+\tilde{u}_2)^2}\Bigg(\frac{1}{3 \pi{\sigma}^{3/2} \left(1+\tilde{u}_2\right)^2}-\frac{\mathrm{F}_\pm^\prime(\tilde{u}_2)}{\sigma^{3/2}} \\
&+T(\bar{v}_4,\tilde{u}_2)-\frac{1}{3\pi\sigma^{3/2}}\frac{1+\frac{3}{5}T(\bar{v}_4,\tilde{u}_2) }{(1+\tilde{u}_2)^2}\Bigg)\,.
\end{align}

\begin{figure}
\begin{center}
\includegraphics[scale=0.5]{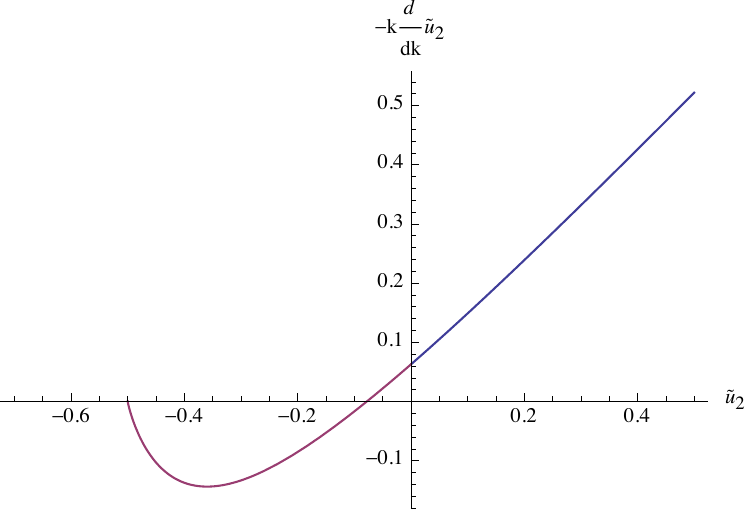}\qquad \includegraphics[scale=0.5]{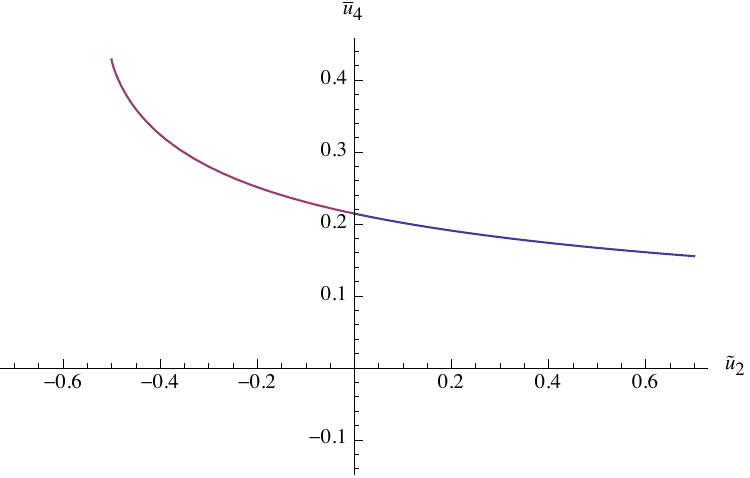}
\end{center}
\caption{Behavior of the $\beta$ function $\dot{\tilde{u}}_2$ (on left) and for $\bar{u}_4(\tilde{u}_2)$ (on right).}\label{figflowDELTA}
\end{figure}

The flow equations are then, in the deep IR:
\begin{align}
\boxed{\dot{\Delta}= -\frac{7\bar{v}_4}{12\sigma^{3/2}}\,(1+\Delta)\, \frac{1-\frac{54}{35}\left(\pi H_{1\pm}(\tilde{u}_2)+ \frac{1}{3\sigma^{3/2}}\frac{1}{(1+\tilde{u}_2)^2} \right)\bar{v}_4}{(1+\tilde{u}_2)^2}\,,}\label{dotDeltaEFF}
\end{align}
and:
\begin{equation}
\boxed{\dot{\tilde{u}}_2=-\tilde{u}_2+\frac{X(\tilde{u}_2,\bar{v}_4,\Delta)}{\frac{1}{3 \pi{\sigma}^{3/2} \left(1+\tilde{u}_2\right)^2}-\frac{\mathrm{F}_\pm^\prime(\tilde{u}_2)}{\sigma^{3/2}} +T(\bar{v}_4,\tilde{u}_2)-\frac{1}{3\pi\sigma^{3/2}}\frac{1+\frac{3}{5}T(\bar{v}_4,\tilde{u}_2) }{(1+\tilde{u}_2)^2}}\,,}\label{flowequ2OOFEFF}
\end{equation}
the coupling $\bar{v}_4$ following the effective flow equation $\dot{\bar{v}}_4=:\beta_{\bar{v}_4}(\bar{v}_4,\tilde{u}_2,\Delta)$. These equations close the hierarchy, and determine, in the local sector, the sextic couplings and beyond. Now, let us investigate these flow equations. First of all, Figure \ref{figflowDELTA} shows the behavior of the flow as $\Delta=v_4=0$. Unsurprising, we recover the results of \cite{lahoche20242} for the equilibrium theory, namely the existence of a repulsive fixed point $\tilde{u}_2^*\approx -0.08$.

\begin{figure}
\begin{center}
\includegraphics[scale=0.7]{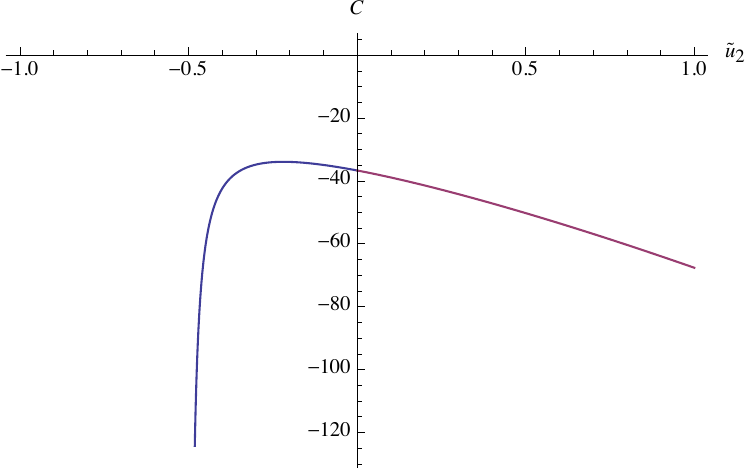}
\end{center}
\caption{Behavior of the function $\mathcal{C}(\tilde{u}_2)$.}\label{figsolutionother}
\end{figure}

Now, let us investigate the other fixed-point solutions. Because of equations \eqref{dotDeltaEFF} and \eqref{equationbetaV4}, we have respectively the two conditions (we set $\sigma=1$):
\begin{equation}
T(\bar{v}_4,\tilde{u}_2)=-\frac{35}{27}\,,
\end{equation}
and, assuming $\Delta \neq -1$:
\begin{equation}
\bar{v}_4(\tilde{u}_2)=\frac{1680}{221}(1+\tilde{u}_2)^2\,.
\end{equation}
Figure \ref{figsolutionother} shows the graphical solution of the equation:
\begin{equation}
\mathcal{C}(\tilde{u}_2):=T\left(\frac{1680}{221}(1+\tilde{u}_2)^2,\tilde{u}_2\right)+\frac{35}{27}\equiv 0\,.
\end{equation}
As the Figure illustrates, there is no additional fixed point for this parametrization of the theory space, and the only one is the global interacting fixed point corresponding to the equilibrium theory:
\begin{equation}
\text{FP}_{\text{Eq}}=\{\tilde{u}_2^*\approx -0.08,\bar{u}_4^*\approx 0.23, \Delta=\bar{v}_4=0\}\,.
\end{equation}
However, the presence of new couplings could change the critical exponents, and by computing them we get:

\begin{equation}
\Theta_{\text{FP}_{\text{Eq}}}=\{\theta_1\approx 0.77,\theta_2\approx 0.69,\theta_3=0\}
\end{equation}

\begin{figure}
\begin{center}
\includegraphics[scale=0.5]{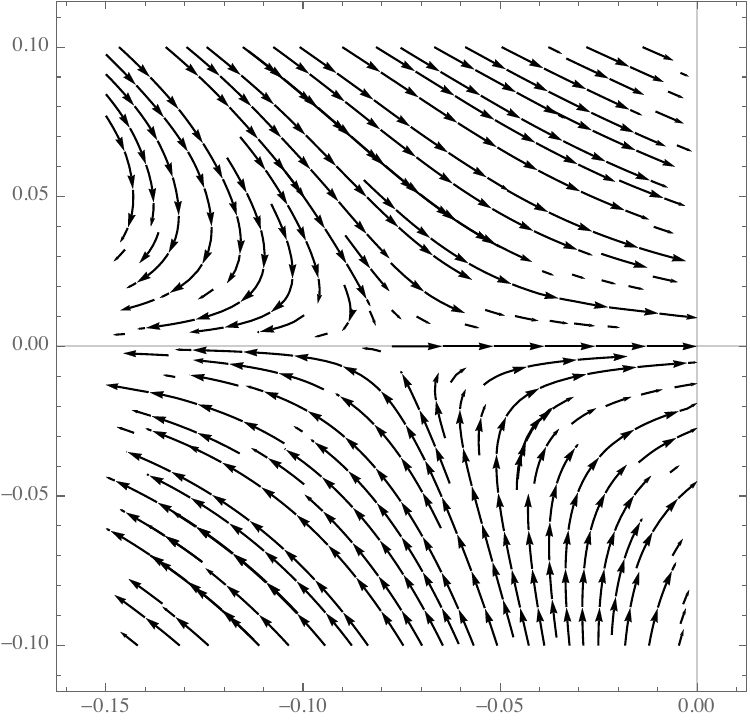}\quad \includegraphics[scale=0.5]{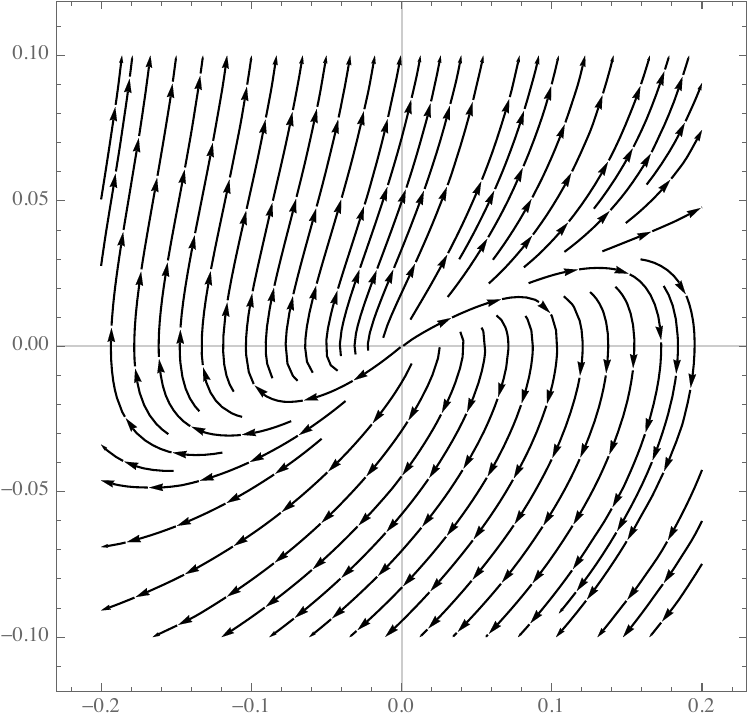}
\end{center}
\caption{Behavior of the renormalization group flow near the fixed point. On the left: in the plane $(\tilde{u}_2,\bar{v}_4)$. On the right: in the plane $(\Delta,\bar{v}_4)$.}\label{Flowdiag}
\end{figure}

We find two relevant and one marginal direction. Furthermore, all the eigenvectors  mixed directions $(\tilde{u}_2,\Delta, \bar{v}_4)$, and none is strictly orthogonal to the plane $(\Delta, \bar{v}_4)$. This fixed point is then unstable in the direction of regions of the theory space where the fluctuation-dissipation theorem does not hold. Figure \ref{Flowdiag} shows the behavior of the renormalization group near the fixed point. These observations suggest that there are no trapped regions of the theory space where the equilibrium phase regime is restored in the deep IR, and the system remains out of equilibrium except if it starts in the equilibrium regime initially. This property is reminiscent of the behavior of the standard $p=2$ spin dynamics, see \cite{de2006random,lahoche2023low,cugliandolo1995full}.

\subsection{Time translation symmetry -- a first look}

In the previous section, we have considered the symmetry breaking of the fluctuation-dissipation theorem, i.e. of time reversal symmetry. Here we will consider a breaking of time translation symmetry, as we have done in \cite{lahoche2023functional} for the $p=2$ soft spin dynamics. We consider the truncation:
 \begin{equation}
\Gamma^{(2)}_k(\vec{p},\omega,\omega^{\prime}) = \begin{pmatrix}
A& B(\vec{p},\omega) \\
B(\vec{p},-\omega)& 0
\end{pmatrix}\delta(\omega-\omega^{\prime})+i\begin{pmatrix}
0& \Delta_k \\
\Delta_k& 0
\end{pmatrix}
\,,\label{truncationTsym}
 \end{equation}
where $\Delta_k$ is a dimensionless constant, independent of $\omega$ and identifying as the order parameter for time translation symmetry. Indeed, at first order (we forget the variable $\vec{p}$ for simplicity):
\begin{equation}
G_{k,\bar{\eta}\Phi}(\omega,\omega^\prime)=B^{-1}(\omega)\delta(\omega-\omega^\prime)-iB^{-1}(\omega)B^{-1}(\omega^\prime) \Delta_k+\mathcal{O}(\Delta_k^2)\,,\label{GLexp}
\end{equation}
Note that this expansion is exact. Indeed, the next order involves the integral $\int \extd \omega B^{-1}(\omega)$. But the heteroclyte condition\footnote{As we see in the beginning of this section, this is also required because the response field does not propagate.} (see Appendix \ref{App2}) imposes in particular that tadpole diagrams contracting both $\bar{\chi}$ and $M$. This condition holds at the bare level because time discretization ensures an additional factor $e^{-i\epsilon \omega}$ (see \cite{canet2011general}), canceling the frequency integral because of the pole position of function $B(\omega)$\footnote{Another argument is that $C_{\bar{\chi} M}(t-t^\prime) \propto \theta(t-t^\prime)$, and that Îto prescription imposes $\theta(0)=0$.}. This simple argument shows that the Ginsburg-Landau expansion \eqref{GLexp} is exact. Using \eqref{eqWARD}, we get:
\begin{equation}
W_{\pm}^{(0)}(k,\tilde{u}_2,0)-\frac{\bar{u}_4}{6}\frac{1}{(1+\tilde{u}_2)^2}+a(\tilde{u}_2) \Delta_k+b(\tilde{u}_2)\Delta^2_k=0\,.
\end{equation}
where $a$ and $b$ are some functions which can be easily computed from \eqref{GLexp} and \eqref{eqWARD}. This constraint determines the quartic local coupling $\bar{u}_4=f(\tilde{u}_2,\Delta_k)$, which allows us to close the pair of flow equations that we can deduce from the truncation \eqref{truncationTsym}. Let us investigate these equations using perturbation in $\Delta_k$ and assuming we are arbitrarily close to the equilibrium dynamics flow. The equation for $\dot{\tilde{u}}_2$ is again given by the equilibrium solution \eqref{flowequ2OOFOS}, and it is easy to check that, at the leading order in $\Delta_k$:
\begin{equation}
\boxed{\dot{\Delta}_k=-\left(\frac{7}{24\sigma^{3/2}}\, \frac{\bar{u}_4^*(\tilde{u}_2)}{(1+\tilde{u}_2)^3}\right) \Delta_k\,.}
\end{equation}
where we used the fact that $\int \extd \omega B^{-1}(\omega)=0$. Because $\bar{u}_4^*(\tilde{u}_2)$ it is positive in the physical domain $\tilde{u}_2>-0.5$ (see Figure \ref{figflowDELTA}), $\dot{\Delta}_k$ is strictly negative and $\Delta_k$ increases toward IR scale. Once again, we recover the instability phenomenon of the equilibrium regime. 

\section{Conclusion and outlooks}\label{sec4}

This paper concludes our investigation series concerning the nonperturbative renormalization group for glassy dynamics of stochastic complex matrices. After the two first papers focused on the equilibrium theory, we have considered the dynamics behavior in the paper, in two different regimes. In the equilibrium dynamics regime, where time-reversal symmetry is expected, the fluctuation-dissipation theorem has to hold along the flow and can be translated as a set of constraints on the truncation. As all these constrained are taken into account, we then showed explicitly that the flow equations and Ward identities reduce to the corresponding expressions computed for equilibrium theory, but with a different interpretation of the different phases (dynamical phase transition). This consistency check is physically expected because equilibrium dynamics means exactly that the system returns at equilibrium for a long time, and we have chosen the initial conditions  $t=-\infty$. In the second time, we investigate our analysis on the out of equilibrium, considering the truncation terms that explicitly violate time-reversal symmetry and then the fluctuation-dissipation theorem. Using Ward identities, we were able to close the hierarchy, and the solution showed the existence of global fixed points controlling the phase transition between equilibrium and out-of-equilibrium dynamics regimes. 
\medskip

Despite all these results, several questions remain open. These issues concern for instance the behavior of the flow beyond the IR limit, a rigorous approach of the deep IR limit, taking into account the "graining" of eigenvalues in the construction of the large $\N$ limit (this especially concern sextic theories, see \cite{lahoche2024letter}), or the improving of the truncated series for in and out of equilibrium regime, in particular regarding the role of higher-order couplings, as pointed out already in \cite{lahoche20242}. Furthermore, the reliability of the approximation scheme used for computing the different terms in the Ward identity should be questioned, and we plan to evaluate it for tensorial models, for which this contribution can be computed exactly in the melonic limit \cite{Lahoche:2018oeo}. Finally, the construction of the truncation  in the out-of-equilibrium regime is more suitable using 2PI rather than 1PI formalism, in the case where the order parameter is the two point function \cite{lahoche2023functional}.

\pagebreak
\appendix
\begin{center}
\begin{LARGE}
\textbf{Additional material}
\end{LARGE}
\end{center}

\section{Martin-Siggia-Rose formalism for complex matrices}\label{App1}

In this section, we provide a short derivation of the MSR path integral for Hermitian matrices. The generating functional can be expressed explicitly as the averaging over the relations of the noise of the functional $\exp (+ L \cdot M_B)$, where $M_B$ is a solution of equation \eqref{eq1} for a fixed $B(t)$:
\begin{equation}
Z_J[L]:=\frac{1}{z_B}\, \int \extd B\, e^{-\frac{1}{2T}\int \extd t\,\Tr\, \eta^2(t)}\, e^{L \cdot M_B}\,,
\end{equation}
with condition $Z_J[L=0]=1$. We use the standard Faddeev–Popov trick and  the formal path integral identity:
\begin{equation}
\int \extd \textbf{q} \, \det \left( \bm{J} \right)\, \prod_{t}\delta\left(\dot{\textbf{q}}(t)+\textbf{f}(t)-\textbf{b}(t)\right)=1\,,\label{identity1}
\end{equation}
where $\bm{J} $ is the matrix with entries:
\begin{equation}
{J}_{\alpha \alpha^\prime}(t-t^\prime)=\frac{\partial}{\partial t}\delta(t-t^\prime) \delta_{\alpha\alpha^\prime} + \frac{\delta^2 \mathcal{H}}{\delta q_{\alpha}(t)\delta q_{\alpha^\prime}(t^\prime)}\,, 
\end{equation}
and we denote by $\textbf{q}:=M(t)$, $\textbf{q}_0:=M(t=0)$, $\bm{q}_x:= \{ \RE(M)_{ij}\}$ and  $\bm{q}_y:= \{ \IM(M)_{ij}\}$
 the $N^2$ vectors corresponding to the real and imaginary parts of the matrix $M$. 
$\bm{x}:= \{ \RE(M)_{ij}\}$ and  $\bm{y}:= \{ \IM(M)_{ij}\}$ the $N^2$ vectors corresponding to the real and imaginary parts of the matrix $M$. In the same way, we denote $\textbf{f}_{ix}$ and $\textbf{f}_{iy}$ the real and imaginary parts of $\textbf{f}_i$, the $i$th component of the $2\N^2$ vector:
\begin{equation}
\textbf{f}=\left\{\RE \left(\frac{\delta \mathcal{H}}{\delta M_{ij}}\right)\,;\IM\left(\frac{\delta \mathcal{H}}{\delta M_{ij}}\right) \right\}\,.\label{expfi}
\end{equation}
Finally, $\bm{b}$ is the $2\N^2$ vector built from the matrix $B$ in the same way. The dissipative Langevin equation \eqref{eq1} being of first order, it admits a unique causal solution. Then using the identity
$
\det \left( \bm{J} \right)= e^{\Tr \log \bm J}\,
$
and due to the fact that the only causal propagator is $\propto \theta(t-t^\prime)$, where $\theta(x)$ is the standard Heaviside function, the exponential reads:
\begin{equation}
\det \left( \bm{J} \right) \sim \exp \left(\theta(0)\int_{-\infty}^{+\infty} \extd t \sum_{\alpha}\frac{\partial^2 H}{\partial q_{\alpha}^2(t)}\right)\,.
\end{equation}
In the  Îto convention for path integral discretization (see \ref{remark2}, references \cite{Zinn-Justin:1989rgp,mannella2012ito} and references therein), we must require that $\theta(0)=0$, then we may choose $\det \left( \bm{J} \right)=1$, and inserting the relation \eqref{identity1} into the functional $Z_J[L]$, we get:
\begin{align}
Z_J^\mathbb{C}[L]&=\frac{1}{z_B}\,\int \extd \textbf{q} \,\int \extd \textbf{b}\, e^{-\int \extd t\frac{\textbf{b}_{x}^2(t)+\textbf{b}_{y}^2(t)}{T}}\, e^{\bar{L} \cdot M_B+L\cdot \bar{M}_B}\, \prod_{t}\delta\left(\dot{\textbf{q}}(t)+\textbf{f}(t)-\textbf{b}(t)\right)\,,\\
&=\frac{1}{z_B}\,\int \extd \textbf{q} \,\left[e^{-\frac{1}{T}\int \extd t (\dot{\textbf{q}}(t)+\textbf{f}(t))_x^2}\times e^{-\frac{1}{T}\int \extd t (\dot{\textbf{q}}(t)+\textbf{f}(t))_y^2}\right]e^{\bar{L} \cdot M+L\cdot \bar{M}}\\
&=\frac{1}{z_B} \,\int \extd \textbf{q} \, e^{-\frac{1}{T}\int \extd t (\dot{\phi}+\delta_{\bar{\psi}}\mathcal{H})\cdot  (\dot{\bar{\phi}}+\delta_{{\psi}}\mathcal{H})}\,,
\end{align}
where $L$ and $\tilde{L}$ are the source fields. The squares can be broken by using the standard Hubbard-Stratonovich trick \cite{moshe2003quantum}. We introduce the pair of $\N\times \N$ complex matrices $\chi$ and $\bar{\chi}$, so that the partition function reads:
\begin{equation}
Z_{J,K}[L,\tilde{L}]:= \int \extd M \extd \overline{M} \extd \chi \extd \overline{\chi}\, e^{- S[M,\overline{M},\chi,\overline{\chi}]+\bar{L}\cdot M+\bar{M}\cdot L + \overline{\tilde{L}}\cdot \chi + \overline{\chi}\cdot \tilde{L}}\,,\label{outofeqbis}
\end{equation}
where the MSR action is:
\begin{equation}
S[M,\bar{M},\chi,\overline{\chi}]:=\int_{-\infty}^{+\infty} \extd t\,\Tr\, \Big[T {\chi}^\dagger(t) \chi(t)+i{\chi}^\dagger(t) \big(\dot{M}+\partial_{\overline{M}} \mathcal{U}\big)+i  \big(\dot{{M}}^\dagger+\partial_{{M}^\tau} \mathcal{U}\big){\chi}(t) \Big]\,.\label{actionMSR2bis}
\end{equation}\,.

\section{Sketched proof of heteroclicity}\label{App2}

In this section, we propose an elementary proof of heteroclicity, based on a causality argument inspired from \cite{canet2011general,lahoche2022stochastic}. We focus on the regulator we consider in this paper, which does not include a coarse-graining over time. For the purpose of this section, we denote it as:
\begin{equation}
R_k(p_1,p_2,t-t^\prime) = Z(2k-p_1-p_2)\theta(k-p_1)\theta(k-p_2)\delta(t-t^\prime)
\end{equation}
including explicitly the Dirac delta ensuring locality in time. We have to prove that:
\begin{equation}
\Gamma_k \Big\vert_{\eta=\bar{\eta}=0}=0\,,\qquad \forall\,k\,.
\end{equation}
Obviously, this holds for $k=\Lambda$ (for some UV cut-off), because $\Gamma_k$ reduces to the MSR classical action which has no term without a response field. We have then to prove that:
\begin{equation}
\frac{\extd}{\extd k}\Gamma_k \Big\vert_{\eta=\bar{\eta}=0}=0\,.
\end{equation}
To check this, we have to notice that, because of the Wetterich flow equation \eqref{Wett2}, the right-hand side involves two terms. The first one involves the product
\begin{equation}
\frac{\extd}{\extd k} R_k(p_1,p_2,t-t^\prime) \langle \bar{\eta}_{\vec{p}}(t) M_{\vec{p}}(t^\prime) \rangle \,,
\end{equation}
but causality must require, $\langle \bar{\eta}_{\vec{p}}(t) M_{\vec{p}}(t^\prime) \rangle\propto \theta(t^\prime-t)$, as it is explicit in \eqref{effectivepropaout}, due to the position of the pole of the Fourier transform. Therefore, because $R_k(p_1,p_2,t-t^\prime) \propto \delta(t-t^\prime)$, the product involves $\theta(0)$, which vanishes because of the Îto prescription.

\printbibliography[heading=bibintoc]

@article{lahoche2024letter,
title={Intriguing connection between Bardeen-Moshe-Bander phenomenon and 2 + p spin
glasses},
author={Lahoche, Vincent and Ousmane Samary, Dine},
doi="10.48550/arXiv.2404.05436",
journal={arXiv:2404.05436 },
year={2024}
}

@article{lahoche20242,
title={Functional renormalization group for “p = 2” like
glassy matrices in the planar approximation: 2 Ward identities method in the deep IR},
author={Lahoche, Vincent and Ousmane Samary, Dine},
journal={arXiv preprint},
doi="arXiv:2403.12217",
year={2024}
}

@article{lahoche20241,
  title={Functional renormalization group for “p = 2” like
glassy matrices in the planar approximation: 1 Vertex expansion at equilibrium},
  author={Lahoche, Vincent and Ousmane Samary, Dine},
  journal={arXiv preprint},
doi="arXiv:2403.07577",
  year={2024}
}

@article{moshe2003quantum,
  title={Quantum field theory in the large N limit: A Review},
  author={Moshe, Moshe and Zinn-Justin, Jean},
  journal={Physics Reports},
doi="10.1016/S0370-1573(03)00263-1",
  volume={385},
  number={3-6},
  pages={69--228},
  year={2003},
  publisher={Elsevier}
}

@book{forrester2010log,
  title={Log-gases and random matrices (LMS-34)},
  author={Forrester, Peter J},
  year={2010},
  publisher={Princeton University Press}
}

@article{Lahoche:2018oeo,
    author = "Lahoche, Vincent and Ousmane Samary, Dine",
    title = "{Nonperturbative renormalization group beyond melonic sector: The Effective Vertex Expansion method for group fields theories}",
    eprint = "1809.00247",
    archivePrefix = "arXiv",
    primaryClass = "hep-th",
    doi = "10.1103/PhysRevD.98.126010",
    journal = "Phys. Rev. D",
    volume = "98",
    number = "12",
    pages = "126010",
    year = "2018"
}

@book{Zinn-Justin:1989rgp,
    author = "Zinn-Justin, Jean",
    title = "{Quantum field theory and critical phenomena}",
    isbn = "978-0-19-850923-3, 978-0-19-883462-5",
    publisher = "Oxford University Press",
    series = "International Series of Monographs on Physics",
doi="10.1093/acprof:oso/9780198509233.001.0001",
    volume = "77",
    month = "4",
    year = "2021"
}

@article{lahoche2022stochastic,
 author = "Lahoche, Vincent and Ousmane Samary, Dine",
    title = "{Stochastic dynamics for group field theories}",
    eprint = "2209.02321",
    archivePrefix = "arXiv",
    primaryClass = "math-ph",
    doi = "10.1103/PhysRevD.107.086009",
    journal = "Phys. Rev. D",
    volume = "107",
    number = "8",
    pages = "086009",
    year = "2023"
}

@article{lahoche2023low,
  title={Low-temperature dynamics for confined $p= 2$ soft spin in the quenched regime},
  author={Lahoche, Vincent and Ousmane Samary, Dine},
  journal={Eur. Phys. J. Plus},
doi="10.1140/epjp/s13360-023-04039-5",
  volume={138},
  number={449},
  pages={1--10},
  year={2023},
  publisher={Springer}
}

@article{lahoche2023functional,
  title={Functional renormalization group for multilinear disordered Langevin dynamics II: Revisiting the p= 2 spin dynamics for Wigner and Wishart ensembles},
  author={Lahoche, Vincent and Ousmane Samary, Dine and Tamaazousti, Mohamed},
  journal={J. Phys. Comm},
doi="10.1088/2399-6528/acd09d",
  year={2023}
}

@article{lahoche2022functional,
  author = "Lahoche, Vincent and Ousmane Samary, Dine  and Tamaazousti, Mohamed",
    title = "{Functional renormalization group for multilinear disordered Langevin dynamics II: Revisiting the $p=2\,$ spin dynamics for Wigner and Wishart ensembles}",
    eprint = "2212.05649",
    archivePrefix = "arXiv",
    primaryClass = "hep-th",
    doi = "10.1088/2399-6528/acd09d",
    journal = "J. Phys. Comm.",
doi="10.1088/2399-6528/acd09d",
    volume = "7",
    number = "5",
    pages = "055005",
    year = "2023"
}

@article{Carrozza_2017,
	doi = {10.1103/physrevd.96.066007},
  
	url = {https://doi.org/10.1103%2Fphysrevd.96.066007},
  
	year = 2017,
	month = {sep},
  
	publisher = {American Physical Society ({APS})},
  
	volume = {96},
  
	number = {6},
  
	author = {Sylvain Carrozza and Vincent Lahoche and Daniele Oriti},
  
	title = {Renormalizable group field theory beyond melonic diagrams: An example in rank four},
  
	journal = {Physical Review D}
}

@article{Carrozza_2017a,
	doi = {10.1088/1361-6382/aa6d90},
  
	url = {https://doi.org/10.1088%2F1361-6382%2Faa6d90},
  
	year = 2017,
	month = {may},
  
	publisher = {{IOP} Publishing},
  
	volume = {34},
  
	number = {11},
  
	pages = {115004},
  
	author = {Sylvain Carrozza and Vincent Lahoche},
  
	title = {Asymptotic safety in three-dimensional {SU}(2) group field theory: evidence in the local potential approximation},
  
	journal = {Classical and Quantum Gravity}
}

@article{Lahoche_2017bb,
    author = "Lahoche, Vincent and Ousmane Samary, Dine",
    title = "{Functional renormalization group for the U(1)-T$_5^6$ tensorial group field theory with closure constraint}",
    eprint = "1608.00379",
    archivePrefix = "arXiv",
    primaryClass = "hep-th",
    doi = "10.1103/PhysRevD.95.045013",
    journal = "Phys. Rev. D",
    volume = "95",
    number = "4",
    pages = "045013",
    year = "2017"
}

@incollection{Delamotte_2012,
	doi = {10.1007/978-3-642-27320-9_2},
  
	url = {https://doi.org/10.1007%2F978-3-642-27320-9_2},
  
	year = 2012,
	publisher = {Springer Berlin Heidelberg},
  
	pages = {49--132},
  
	author = {Bertrand Delamotte},
  
	title = {An Introduction to the Nonperturbative Renormalization Group},
  
	booktitle = {Renormalization Group and Effective Field Theory Approaches to Many-Body Systems}
}

@article{benedetti2016functional,
  title={Functional renormalization group approach for tensorial group field theory: a rank-6 model with closure constraint},
  author={Benedetti, Dario and Lahoche, Vincent},
  journal={Classical and Quantum Gravity},
doi="10.1088/0264-9381/33/9/095003",
  volume={33},
  number={9},
  pages={095003},
  year={2016},
  publisher={IOP Publishing}
}

@article{canet2011general,
    author = "Canet, Leonie and Chate, Hugues and Delamotte, Bertrand",
    title = "{General framework of the non-perturbative renormalization group for non-equilibrium steady states}",
    eprint = "1106.4129",
    archivePrefix = "arXiv",
    primaryClass = "cond-mat.stat-mech",
    doi = "10.1088/1751-8113/44/49/495001",
    journal = "J. Phys. A",
    volume = "44",
    pages = "495001",
    year = "2011"
}

@article{cugliandolo1995full,
  title={Full dynamical solution for a spherical spin-glass model},
  author={Cugliandolo, Leticia F and Dean, David S},
  journal={Journal of Physics A: Mathematical and General},
doi="10.1088/0305-4470/28/15/003",
  volume={28},
  number={15},
  pages={4213},
  year={1995},
  publisher={IOP Publishing}
}

@article{aron2010symmetries2,
    author = "Aron, Camille and Biroli, Giulio and Cugliandolo, Leticia F.",
    title = "{Symmetries of generating functionals of Langevin processes with colored multiplicative noise}",
    eprint = "1007.5059",
    archivePrefix = "arXiv",
    primaryClass = "cond-mat.stat-mech",
    doi = "10.1088/1742-5468/2010/11/P11018",
    journal = "J. Stat. Mech.",
    volume = "1011",
    pages = "P11018",
    year = "2010"
}

@article{duclut2017frequency,
    author = "Duclut, Charlie and Delamotte, Bertrand",
    title = "{Frequency regulators for the nonperturbative renormalization group: A general study and the model A as a benchmark}",
    eprint = "1611.07301",
    archivePrefix = "arXiv",
    primaryClass = "cond-mat.stat-mech",
    doi = "10.1103/PhysRevE.95.012107",
    journal = "Phys. Rev. E",
    volume = "95",
    number = "1",
    pages = "012107",
    year = "2017"
}

@article{lahoche2021functional,
    author = "Lahoche, Vincent and Ousmane Samary, Dine and Ouerfelli, Mohamed",
    title = "{Functional renormalization group for multilinear disordered Langevin dynamics I: Formalism and first numerical investigations at equilibrium}",
    eprint = "2106.05690",
    archivePrefix = "arXiv",
    primaryClass = "hep-th",
    doi = "10.1088/2399-6528/ac61b3",
    journal = "J. Phys. Comm.",
    volume = "6",
    pages = "055002",
    year = "2022"
}

@book{potters2020first,
  title={A First Course in Random Matrix Theory: For Physicists, Engineers and Data Scientists},
  author={Potters, Marc and Bouchaud, Jean-Philippe},
  year={2020},
  publisher={Cambridge University Press},
doi="10.1017/9781108768900"
}

@book{de2006random,
  title={Random fields and spin glasses: a field theory approach},
  author={De Dominicis, Cirano and Giardina, Irene},
  year={2006},
  publisher={Cambridge University Press},
doi="10.1017/CBO9780511534836"
}

@article{kleinert2011hubbard,
  title={Hubbard-Stratonovich transformation: Successes, failure, and cure},
  author={Kleinert, Hagen},
doi="10.48550/arXiv.1104.5161",
  journal={arXiv preprint arXiv:1104.5161},
  year={2011}
}

@article{martin1973statistical,
  title={Statistical dynamics of classical systems},
  author={Martin, Paul Cecil and Siggia, ED and Rose, HA},
  journal={Physical Review A},
doi="10.1103/PhysRevA.8.423",
  volume={8},
  number={1},
  pages={423},
  year={1973},
  publisher={APS}
}

@article{castellani2005spin,
  title={Spin-glass theory for pedestrians},
  author={Castellani, Tommaso and Cavagna, Andrea},
  journal={Journal of Statistical Mechanics: Theory and Experiment},
  volume={2005},
  number={05},
  pages={P05012},
  year={2005},
  publisher={IOP Publishing},
doi="10.1088/1742-5468/2005/05/P05012"
}

@article{mannella2012ito,
  title={It{\^o} versus Stratonovich: 30 years later},
  author={Mannella, Riccardo and McClintock, Peter VE},
  journal={Fluctuation and Noise Letters},
doi="10.1142/S021947751240010X",
  volume={11},
  number={01},
  pages={1240010},
  year={2012},
  publisher={World Scientific}
}

@article{bouchaud1996mode,
  title={Mode-coupling approximations, glass theory and disordered systems},
  author={Bouchaud, Jean-Philippe and Cugliandolo, Leticia and Kurchan, Jorge and M{\'e}zard, Marc},
  journal={Physica A: Statistical Mechanics and its Applications},
doi="10.1016/0378-4371(95)00423-8",
  volume={226},
  number={3-4},
  pages={243--273},
  year={1996},
  publisher={Elsevier}
}
\end{document}